\documentclass[11pt]{article}
\usepackage{amsmath, amsfonts, amssymb, times, fullpage}

\usepackage{ifthen}
\usepackage{amsmath}
\usepackage{amssymb}
\usepackage{amscd}
\usepackage{graphicx}
\usepackage{latexsym}
\usepackage{psfig}
\usepackage{here}
\usepackage{color}

\usepackage{fullpage}
\usepackage[lined, linesnumbered, noresetcount, boxruled]{algorithm2e}
\usepackage{multicol}

\usepackage{subfig}

\newcommand{\remove}[1]{}

\newcommand{\lref}[1]{line~\nlSty{\ref{#1}}}
\newcommand{\Lref}[1]{Line~\nlSty{\ref{#1}}}
\newcommand{\llref}[2]{lines~\nlSty{\ref{#1}}--\nlSty{\ref{#2}}}
\newcommand{\Llref}[2]{Lines~\nlSty{\ref{#1}}--\nlSty{\ref{#2}}}

\newtheorem{definition}{Definition}

\newtheorem{lemma}{Lemma}

\newtheorem{theorem}{Theorem}

\newcommand{\qedsymb}{\hfill{\rule{2mm}{2mm}}}

\newenvironment{proofsketch}{\begin{trivlist}
\item[\hspace{\labelsep}{\bf\noindent Proof sketch: }]
}{\qedsymb\end{trivlist}}

\begin{document}

%\begin{titlepage}
\begin{titlepage}
\title{A Dynamic Elimination-Combining Stack Algorithm}

\author{Gal Bar-Nissan, Danny Hendler and Adi Suissa \\
        Department of Computer Science \\
        Ben-Gurion University}

\maketitle
\thispagestyle{empty}

\begin{abstract}

Two key synchronization paradigms for the construction of scalable concurrent data-structures are \emph{software combining} and \emph{elimination}. Elimination-based concurrent data-structures allow operations with \emph{reverse semantics} (such as \emph{push} and \emph{pop} stack operations) to ``collide'' and exchange values without having to access a central location. Software combining, on the other hand, is effective when colliding operations have \emph{identical semantics}: when a pair of threads performing operations with identical semantics collide, the task of performing the combined set of operations is delegated to one of the threads and the other thread waits for its operation(s) to be performed. Applying this mechanism iteratively can reduce memory contention and increase throughput.

The most highly scalable prior concurrent stack algorithm is the \emph{elimination-backoff stack} \cite{DBLP:journals/jpdc/HendlerSY10}. The elimination-backoff stack provides high parallelism for symmetric workloads in which the numbers of \emph{push} and \emph{pop} operations are roughly equal, but its performance deteriorates when workloads are asymmetric.

We present DECS, a novel Dynamic Elimination-Combining Stack algorithm, that scales well for all workload types. While maintaining the simplicity and low-overhead of the elimination-bakcoff stack, DECS manages to benefit from collisions of both identical- and reverse-semantics operations. Our empirical evaluation shows that DECS scales significantly better than both blocking and non-blocking best prior stack algorithms.\\
\vspace{1cm}

\end{abstract}
\end{titlepage}
\pdfoutput=1 
%\end{titlepage}
%\renewcommand{\baselinestretch}{1.1}\normalsize

%\newcounter{linenumber}
%\def\rrnnll{\setcounter{linenumber}{0}}
%\def\nnll{\refstepcounter{linenumber}\lf\thelinenumber}

%****************************intro.tex**********************************

\section{Introduction}
Concurrent stacks are widely used in parallel applications and
operating systems. As shown in \cite{taura93efficient}, LIFO-based
scheduling reduces excessive task creation and prevents threads from attempting
to dequeue and execute a task which depends on the results of other tasks. A concurrent
stack supports the \emph{push} and \emph{pop} operations with linearizable LIFO semantics.
\textit{Linearizability} \cite{HerlihyWingLinearizability}, which is the most widely used correctness condition for concurrent objects, guarantees that each operation appears to have an atomic effect at some point between its invocation and response and that operations can be combined in a modular way.

Two key synchronization paradigms for the construction of scalable concurrent data-structures in general, and concurrent stacks in particular, are \emph{software combining} \cite{YewTL87,DBLP:journals/dc/HendlerK09,DBLP:conf/opodis/HendlerKM10} and \emph{elimination} \cite{DBLP:conf/europar/AfekKNS10,touitou97eltree}. Elimination-based concurrent data-structures allow operations with reverse semantics (such as \emph{push} and \emph{pop} stack operations) to ``collide'' and exchange values without having to access a central location. Software combining, on the other hand, is effective when colliding operations have identical semantics: when a pair of threads performing operations with identical semantics collide, the task of performing the combined set of operations is delegated to one of the threads and the other thread waits for its operation(s) to be performed. Applying this mechanism iteratively can reduce memory contention and increase throughput.

The design of efficient stack algorithms poses several challenges. Threads sharing the stack implementation must synchronize to ensure correct linearizable executions. To provide scalability, a stack algorithm must be highly parallel; this means that, under high load, threads must be able to synchronize their operations without accessing a central location in order to avoid sequential bottlenecks. Scalability at high loads should not, however, come at the price of good performance in the more common low contention cases. Hence, another challenge faced by stack algorithms is to ensure low latency of stack operations when only a few threads access the stack simultaneously.

The most highly scalable concurrent stack algorithm known to date is the lock-free \emph{elimination-backoff stack} of Hendler, Shavit and Yerushalmi \cite{DBLP:journals/jpdc/HendlerSY10} (henceforth referred to as the HSY stack). It uses a single elimination array as a backoff scheme on a simple lock-free central stack (such as Treiber's stack algorithm \cite{treiber}\footnote{Treiber's algorithm is a variant of an algorithm previously introduced by IBM \cite{IBM370-manual}.}). If the threads fail on the central stack, they attempt to eliminate on the array, and if they fail in eliminating, they attempt to access the central stack once again and so on. As shown by Michael and Scott \cite{michael98nonblocking}, the central stack of \cite{treiber} is highly efficient under low contention. Since threads use the elimination array only when they fail on the central stack, the elimination-backoff stack algorithm enjoys similar low contention efficiency.

The HSY stack scales well under high contention if the workload is symmetric (that is, the numbers of \emph{push} and \emph{pop} operations are roughly equal), since multiple pairs of operations with reverse semantics succeed in exchanging values without having to access the central stack. Unfortunately, when workloads are asymmetric, most collisions on the elimination array are between operations with identical semantics. For such workloads, the performance of the HSY stack deteriorates and falls back to the sequential performance of a central stack.

Recent work by Hendler et al. introduced \emph{flat-combining} \cite{FlatCombining}, a synchronization mechanism based on coarse-grained locking in which a single thread holding a lock performs the combined work of other threads. They presented flat-combining based implementations of several concurrent objects, including a flat-combining stack (FC stack). Due to the very low synchronization overhead of flat-combining, the FC stack significantly outperforms other stack implementations (including the elimination-backoff stack) in low and medium concurrency levels. However, since the FC stack is essentially sequential, its performance does not scale and even deteriorates when concurrency levels are high.

\remove{
Two known methods for parallelizing shared stacks do not meet the aforementioned design goals.
The combining funnels of Shavit and Zemach \cite{funnels} are blocking linearizable LIFO stacks that offer
scalability through combining, but perform poorly at low loads
because of the combining overhead. The elimination trees of Shavit and Touitou \cite{touitou97eltree} are
non-blocking \cite{waitfree}, but the stack they provide is not
linearizable, and it too has large overheads that cause it to
perform poorly at low loads. On the other hand, the results of
Michael and Scott \cite{michael98nonblocking} show that the best
known low load method, the simple linearizable lock-free stack introduced by IBM \cite{IBM370-manual} (a variant of which was later presented by Treiber \cite{treiber}), scales poorly due to contention and an
inherent sequential bottleneck.
} %%% END REMOVE - possibly move to discussion

\subsubsection*{Our Contributions:}

This paper presents DECS, a novel Dynamic Elimination-Combining Stack algorithm, that scales
well for all workload types. While maintaining the simplicity and low-overhead of the
HSY stack, DECS manages to benefit from collisions of both identical- and reverse-semantics operations.

\begin{figure}[b]
\begin{center}
\includegraphics[width=120mm]{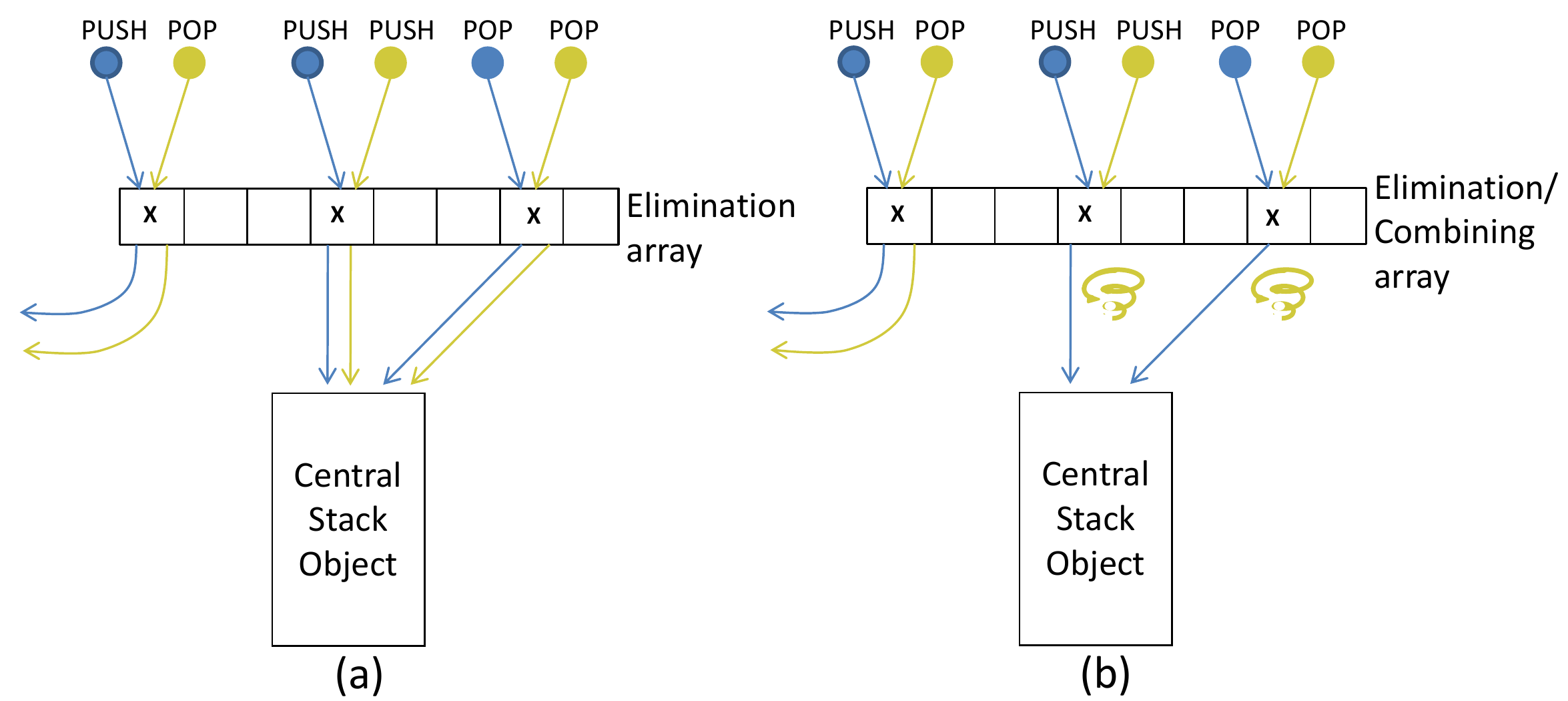}
\end{center}
\caption{Collision-attempt scenarios: (a) Collision scenarios in the elimination-backoff stack; (b) Collision scenarios in DECS.} \label{figure:collision-scenarios}
\end{figure}

The idea underlying DECS is simple. Similarly to the HSY stack, DECS uses a contention-reduction layer as a backoff scheme for a central stack. However, whereas the HSY algorithm uses an elimination layer, DECS uses an \emph{elimination-combining layer} on which concurrent operations can dynamically either eliminate or combine, depending on whether their operations have reverse or identical semantics, respectively. As illustrated by Figure \ref{figure:collision-scenarios}-(a), when two identical-semantics operations executing the HSY algorithm collide, both have to retry their operations on the central stack. With DECS (Figure \ref{figure:collision-scenarios}-(b)), every collision, regardless of the types of the colliding operations, reduces contention on the central stack and increases parallelism by using either elimination or combining. Since combining is applied iteratively, each colliding operation may attempt to apply the combined operations (multiple \emph{push} or multiple \emph{pop} operations) of multiple threads - its own and (possibly) the operations delegated to it by threads with which it previously collided, threads that are awaiting their response.

We compared DECS with a few prior stack algorithm, including the HSY and the FC stacks. DECS outperforms the HSY stack on all workload types and all concurrency levels; specifically, for asymmetric workloads, DECS provides up to 3 times the throughput of the HSY stack.

The FC stack outperforms DECS in low and medium levels of concurrency. The performance of the FC stack deteriorates quickly, however, as the level of concurrency increases. DECS, on the other hand, continues to scale on all workload types and outperforms the FC stack in high concurrency levels by a wide margin, providing up to $10$ times its throughput.

For some applications, a nonblocking \cite{waitfree} stack may be preferable to a blocking one because lock-freedom is more robust in the face of thread failures. Whereas the elimination-backoff stack is lock-free, both the FC and the DECS stacks are blocking. We present NB-DECS, a lock-free \cite{waitfree} variant of DECS that allows threads that delegated their operations to a combining thread and have waited for too long to cancel their ``combining contracts'' and retry their operations. The performance of NB-DECS is slightly better than that of the HSY stack when workloads are symmetric and for \emph{pop}-dominated workloads, but it provides significantly higher throughput for \emph{push-dominated} asymmetric workloads.

The remainder of this paper is organized as follows. We describe the DECS algorithm in Section \ref{sec:algorithm} and report on its performance evaluation in Section \ref{sec:evaluation}. A high-level description of the NB-DECS algorithm and its performance evaluation are provided in Section \ref{sec:NB-DECS}. We conclude the paper in Section \ref{sec:discussion} with a short discussion of our results. Detailed correctness proof sketches and pseudo-codes of some functions that were omitted from the body of the paper are presented in the appendix.

%\IncludeFigure{Reverse_Sem}{1.0}{Execution of operations with
%reverse semantics on a stack}{rev_sem}

\section{The Dynamic Elimination-Combining Algorithm}
\label{sec:algorithm}

% Various definitions for the code
\SetKwBlock{Loop}{loop}{end loop}
\SetKw{NCS}{Non Critical Section}
\SetKw{CS}{Critical Section}
%\SetKw{Struct}{struct}
\SetKwData{Int}{\textbf{int}}
\SetKwData{Void}{\textbf{void}}
\SetKwData{Bool}{\textbf{boolean}}
\SetKwData{multiOp}{\textbf{multiOp}}
\SetKwData{Stack}{\textbf{Stack}}
\SetKwData{Cell}{\textbf{Cell}}
\SetKwData{Data}{\textbf{Data}}
\SetKw{Array}{array}
\SetKw{Of}{of}
\SetKw{Define}{define}
\SetKw{Init}{init}
\SetKw{Goto}{goto}
\SetKw{Await}{await}
\SetKw{True}{true}
\SetKw{False}{false}
\SetKw{Null}{null}
\SetKw{Cont}{continue loop}
\SetKw{Local}{local:}
\SetKw{Shared}{shared}
\SetKw{Global}{global}
\SetKw{Constant}{constant}
\SetKw{AndCond}{$\land$}
\SetKw{OrCond}{$\lor$}
\newcommand{\NIL}{\ensuremath{\bot}}
% End various definitions for the code

In this section we describe the DECS algorithm. Figure \ref{fig:DS-centralops-collide}-(a) presents the data-structures and shared variables used by DECS. Similarly to the HSY stack, DECS uses two global arrays - \emph{location} and \emph{collision} - which comprise its elimination-combining layer. Each entry of the \emph{location} array corresponds to a thread $t \in \{1..N\}$ and is either \emph{NULL} or stores a pointer to a \emph{multiOp} structure described shortly. Each non-empty entry of the \emph{collision} array stores the ID of a thread waiting for another thread to collide with it. DECS also uses a \emph{CentralStack} structure, which is a singly-linked-list of \emph{Cell} structures - each comprising an opaque \emph{data} field and a \emph{next} pointer. Threads iteratively do the following: first, they try to apply their operation to the \emph{CentralStack} structure and, if they fail, they access the elimination-combining layer (see Figure \ref{fig:mainops} in the appendix\footnote{For lack of space, some of the more straightforward pseudo-code is presented in the appendix.}).

\emph{Push} or \emph{pop} operations that access the elimination-combining layer may be combined. Thus, in general, operations that are applied to the central stack or to the elimination-combining layer are \emph{multi-ops}; that is, they are either \emph{multi-pop} or \textit{multi-push} operations which represent the combination of multiple \emph{pop} or \emph{push} operations, respectively. A multi-op is performed by a \emph{delegate thread}, attempting to perform its own operation and (possibly) also those of one or more \emph{waiting threads}. The \emph{length} of a multi-op is the number of operations it consists of (which is the number of corresponding waiting threads plus 1). Each multi-op is represented by a \emph{multiOp} structure (see Figure \ref{fig:DS-centralops-collide}-(a)), consisting of a thread identifier \emph{id}, the operations type (\emph{PUSH} or \emph{POP}) and a \emph{Cell} structure (containing the thread data  in case of a multi-push or empty in case of a multi-pop). The \emph{next} field points to the structure of the next operation of the multiOp (if any). Thus, each multiOp is represented by a \emph{multiOp list} of structures, the first of which represents the operation of the delegate thread. The \emph{last} field points to the last structure in the multiOp list and the \emph{length} field stores the multi-op's length. The \emph{cStatus} field is used for synchronization between a delegate thread and the threads awaiting it and is described later in this section.

\subsubsection*{Central Stack Functions}
\label{subsec:operations}

Figures \ref{fig:DS-centralops-collide}-(b) and \ref{fig:DS-centralops-collide}-(c) respectively present the pseudo-code of the \texttt{cMultiPop} and \texttt{cMultiPush} functions applied to the central stack.

The \texttt{cMultiPop} function receives as its input a pointer to the first \emph{multiOp} record in a multi-op list of \emph{pop} operations to be applied to the central stack. It first reads the central stack pointer (\lref{CPOP:InitTop}). If the stack is empty, then all the \emph{pop} operations in the list are linearized in \lref{CPOP:InitTop} and will return an \emph{empty} indication. In \llref{CPOP:Release}{CPOP:EndCheckTop}, an EMPTY\_CELL is assigned as the response of all these operations and the \emph{cStatus} fields of all the multiOp structures is set to FINISHED in order to signal all waiting threads that their response is ready. The \texttt{cMultiPop} function then returns \emph{true} indicating to the delegate thread that its operation was applied.

If the stack is not empty, the number $m$ of items that should be popped from the central stack is computed (\llref{CPOP:InitNext}{CPOP:EndNextLoop}); this is the minimum between the length of the multi-pop operation and the central stack's size. The \emph{nTop} pointer is set accordingly and a \emph{CAS} is applied to the central stack attempting to atomically pop $m$ items (\lref{CPOP:IfPopCAS}).
If the CAS fails, \emph{false} is returned (\lref{CPOP:ReturnFalse}) indicating that \texttt{cMultiPop} failed and that the multi-pop should be next applied to the elimination-combining layer.

If the CAS succeeds, then all the multi-pop operations are linearized when it occurs. The \texttt{cMultiPop} function proceeds by iterating over the multi-op list (\llref{CPOP:BeginCheckNextLoop}{CPOP:EndCheckNextLoop}). It assigns the $m$ cells that were popped from the central stack to the first $m$ pop operations  (\lref{CPOP:WritingCell}) and assigns an EMPTY\_CELL to the rest of the pop operations, if any (\lref{CPOP:WritingEmptyCell}).  It then sets the \emph{cStatus} of all these operations to FINISHED (\lref{CPOP:ReleasingTinfoNext}), signalling all waiting threads that their response is ready. The \texttt{cMultiPop} function then returns \emph{true}, indicating that it was successful (\lref{CPOP:ReturnTrue}).

The \texttt{cMultiPush} function receives as its input a pointer to the first \emph{multiOp} record in a multi-op list of \emph{push} operations to be applied to the central stack. It sets the \emph{next} pointer of the last cell to point to the top of the central stack (\lref{CPUSH:setlast}) and applies a \textit{CAS} operation in an attempt to atomically chain the list to the central stack (\lref{CPUSH:castop}). If the CAS succeeds, then all the \emph{push} operations in the list are linearized when it occurs. In this case, the \texttt{cMultiPush} function proceeds by iterating over the multi-op list and setting the \emph{cStatus} of the \emph{push} operations to FINISHED (\llref{CPUSH:startWhile}{CPUSH:endWhile}). It then returns \emph{true} in \lref{CPUSH:returnTrue}, indicating its success. If the CAS fails, \texttt{cMultiPush} returns \emph{false} (\lref{CPUSH:returnFalse}) indicating that the multi-push should now be applied to the elimination-combining layer.

\subsubsection*{Elimination-Combining Layer Functions}

The \texttt{collide} function, presented in Figure \ref{fig:DS-centralops-collide}-(d), implements the elimination-combining backoff algorithm performed after a multi-op fails on the central stack.\footnote{This function is similar to the \texttt{LesOP} function of the HSY stack and is described for the sake of presentation completeness.} It receives as its input a pointer to the first \emph{multiOp} record in a multi-op list. A delegate thread executing the function first \textit{registers} by writing to its entry in the \emph{location} array (\lref{COLLIDE:announce}) a pointer to its \emph{multiOp} structure, thus advertising itself to other threads that may access the elimination-combining layer . It then chooses randomly and uniformly an index into the collision array (\lref{COLLIDE:getpos}) and repeatedly attempts to swap the value in the corresponding entry with its own ID by using \emph{CAS} (\llref{COLLIDE:gethim}{COLLIDE:gethimagain}).

%We now describe the DECS algorithm code.
%Figure \ref{fig:mainops} shows the push and pop operations code that can be executed by a thread.
%The code in figure \ref{fig:centralops} follows Treiber's \cite{treiber} algorithm and describes the stack algorithm %on the central stack.
%The \emph{collide} function described in figure \ref{fig:collide} is similar to the HSY stack \emph{LesOP} function, %and is used to combine or eliminate two threads' multi-operations.
%Figure \ref{fig:combineEliminate} describes the DECS elimination and combining operations.
%A thread starts by attempting to execute a pop (push) operation on the central stack.
%If this attempt is unsuccessful, the thread then executes the collision procedure.
%A collision occurs between two threads that execute the collision procedure, called an active thread and a passive %thread.
%An \emph{active thread} is the thread that attempts to collide with a passive thread.
%A \emph{passive thread} is a thread that awaits for some active thread to collide with it.
%When both threads have the same operation semantics, these operations are eliminated.
%If the semantics are reversed, then the passive thread delegates its operation to the active thread, and waits until %the active thread executes the operation.

\begin{figure}[H]
\caption[Central Stack Operations]{(a): Data structures, (b), (c): central stack operations, (d): the \texttt{collide} function.}
\label{fig:DS-centralops-collide}

\begin{multicols}{2}
%\NoCaptionOfAlgo
\nocaptionofalgo

\begin{algorithm}[H]
%\SetAlgoLined
\SetLine
%%%%%%%%%%%%%%%%%%%%%%%%%%%%%%%%%%%%%%%%
\caption{(a) Data Structures and Shared Variables}
%%%%%%%%%%%%%%%%%%%%%%%%%%%%%%%%%%%%%%%%
\Define  Cell: struct \{\emph{data}: \Data, \emph{next}: \Cell\}\;
\Define  multiOp: struct
\{\emph{id},\emph{op},\emph{length},\emph{cStatus}:\Int, \emph{cell}: \Cell,
\emph{next},\emph{last}: \multiOp\}\;
%\Constant \emph{EMPTY\_CELL}: \Cell \Init \{ \Null, \Null\}\;
\Global \emph{CentralStack}: \Cell \;
\Global \emph{collision}: array of $[$1,\ldots,N$]$ of \Int \Init \emph{EMPTY}\;
\Global \emph{location}: array of $[$1,\ldots,N$]$ of \multiOp \Init \Null\;

\end{algorithm}

\begin{algorithm}[H]
%\SetAlgoLined
\SetLine
%%%%%%%%%%%%%%%%%%%%%%%%%%%%%%%%%%%%%%%%
\caption{(b) \textbf{boolean} \texttt{cMultiPop}(\emph{multiOp}: mOp)}
%%%%%%%%%%%%%%%%%%%%%%%%%%%%%%%%%%%%%%%%
    \emph{top} = \emph{CentralStack}\nllabel{CPOP:InitTop}\;
    \If{top = \Null}{ \nllabel{CPOP:CheckTop}
        \Repeat {mOp = \Null}
        {
        \emph{mOp}.\emph{cell} = EMPTY\_CELL\;\nllabel{CPOP:Release}
        \emph{mOp}.\emph{cStatus} = FINISHED\;\nllabel{CPOP:ReleaseAll}
        \emph{mOp}=\emph{mOp}.\emph{next}\;
        }
        \Return \True\;\nllabel{CPOP:ReturnEmpty}
    } \nllabel{CPOP:EndCheckTop}
    \emph{nTop} = \emph{top}.\emph{next}\nllabel{CPOP:InitNext}\;
	\emph{m} = 1\nllabel{CPOP:InitnumToPop}\;
    \While{nTop $\neq$ \Null \AndCond m $<$ mOp.length}{\nllabel{CPOP:BeginNextLoop}
        \emph{nTop} = \emph{nTop}.\emph{next}\nllabel{CPOP:IncrementNext},  \emph{m}++\nllabel{CPOP:IncrementCurSize}\;
    }\nllabel{CPOP:EndNextLoop}
    \uIf{CAS(\&CentralStack, top, nTop)}{ \nllabel{CPOP:IfPopCAS}
        \emph{mOp}.\emph{cell} = \emph{top}\nllabel{CPOP:PoppingTinfo}\;
        \emph{top} = \emph{top}.\emph{next}\nllabel{CPOP:IncrementingTop}\;
        \While{mOp.next $\neq$ \Null}{ \nllabel{CPOP:BeginCheckNextLoop}
            \eIf{top = \Null}{ \nllabel{CPOP:IfTopCellNull}
                \emph{mOp}.\emph{next}.\emph{cell} = EMPTY\_CELL\nllabel{CPOP:WritingEmptyCell}\;
            }{\nllabel{CPOP:ElseTopCellNull}
                \emph{mOp}.\emph{next}.\emph{cell} = top\nllabel{CPOP:WritingCell}\;
                \emph{top} = \emph{top}.\emph{next}\nllabel{CPOP:IncrementingCell}\;
            }
            \emph{mOp}.\emph{next}.\emph{cStatus} = FINISHED\nllabel{CPOP:ReleasingTinfoNext}\;
            \emph{mOp}.\emph{next} = \emph{mOp}.\emph{next}.\emph{next}\nllabel{CPOP:IncrementingTinfoNext}\;
        }\nllabel{CPOP:EndCheckNextLoop}
        \Return \True\;\nllabel{CPOP:ReturnTrue}
    }
    \lElse{\nllabel{CPOP:ElsePopCAS}
        \Return \False\;\nllabel{CPOP:ReturnFalse}
    }\nllabel{CPOP:EndPopCAS}
\end{algorithm}

\columnbreak

%\NoCaptionOfAlgo
\nocaptionofalgo
\begin{algorithm}[H]
%\SetAlgoLined
\SetLine
%%%%%%%%%%%%%%%%%%%%%%%%%%%%%%%%%%%%%%%%
\caption{(c) \textbf{boolean} \texttt{cMultiPush}(\textit{multiOp}: mOp)}
%%%%%%%%%%%%%%%%%%%%%%%%%%%%%%%%%%%%%%%%
    \emph{top} = \emph{CentralStack}\;
    \emph{mOp}.\emph{last}.\emph{cell}.\emph{next}=\emph{top}\nllabel{CPUSH:setlast}\;
    \eIf{CAS(\&CentralStack, top, mOp.cell)}{\nllabel{CPUSH:castop}
        \While{mOp.next $\neq$ \Null}{\nllabel{CPUSH:startWhile}
            \emph{mOp}.\emph{next}.\emph{cStatus} = FINISHED\nllabel{CPUSH:setFinish}\;
            \emph{mOp}.\emph{next} = \emph{mOp}.\emph{next}.\emph{next}\;
        }\nllabel{CPUSH:endWhile}
        \Return \True\;\nllabel{CPUSH:returnTrue}
    }{
        \Return \False\;\nllabel{CPUSH:returnFalse}
    }
\end{algorithm}

\begin{algorithm}[H]
%\SetAlgoLined
\SetLine
%%%%%%%%%%%%%%%%%%%%%%%%%%%%%%%%%%%%%%%%
\caption{(d) \textbf{boolean} \texttt{collide}(multiOp: mOp)}
%%%%%%%%%%%%%%%%%%%%%%%%%%%%%%%%%%%%%%%%
    \emph{location}$[$\emph{id}$]$ = \emph{mOp}\nllabel{COLLIDE:announce}\;
    \emph{index} = randomIndex()\nllabel{COLLIDE:getpos}\;
    \emph{him} = \emph{collision}$[$\emph{index}$]$\nllabel{COLLIDE:gethim}\;
    \While{CAS(\&collision$[$index$]$, him, id)=\False \nllabel{COLLIDE:cascollision}}{
        \emph{him} = \emph{collision}$[$\emph{index}$]$\nllabel{COLLIDE:gethimagain}\;
    }
    \If{him $\neq$ EMPTY \nllabel{COLLIDE:checkhim}}{
        \emph{oInfo} = \emph{location}$[$\emph{him}$]$\nllabel{COLLIDE:getinfo}\;
        \If{oInfo $\neq$ NULL \AndCond oInfo.id $\neq$ id \AndCond oInfo.id=him \nllabel{COLLIDE:checkinfo}}{
            \uIf{CAS(\&location$[$id$]$, mOp, NULL)=\True \nllabel{COLLIDE:caslocation1}}{
                \Return \texttt{activeCollide}(\emph{mOp}, \emph{oInfo})\nllabel{COLLIDE:caslocation1success}\;
            }\Else{
                \Return \texttt{passiveCollide}(\emph{mOp})\nllabel{COLLIDE:caslocation1fail}\;
            }
        }
    }
    wait()\nllabel{COLLIDE:delay}\;
    \If{CAS(\&location$[$id$]$, mOp, NULL)=\False \nllabel{COLLIDE:caslocation2}}{
        \Return \texttt{passiveCollide}(mOp)\nllabel{COLLIDE:caslocation2fail}\;
    }\nllabel{COLLIDE:caslocation2end}
    \Return \False\;\nllabel{COLLIDE:return}

\end{algorithm}

\end{multicols}
\end{figure}

A thread that initiates a collision is called an \emph{active collider} and a thread that discovers it was collided with is called a \emph{passive collider}. If the value read from the collision array entry is not null (\lref{COLLIDE:checkhim}), then it is a value written there by another registered thread that may await a collision. The delegate thread (now acting as an active collider) proceeds by reading a pointer to the other thread's multiOp structure \emph{oInfo} (\lref{COLLIDE:getinfo}) and then verifies that the other thread may still be collided with (\lref{COLLIDE:checkinfo}).\footnote{Some of the tests of \lref{COLLIDE:checkinfo} are required because \textit{location} array entries are not re-initialized when operations terminate (for optimization reasons) and thus may contain outdated values.}

If the tests of \lref{COLLIDE:checkinfo} succeed, the delegate thread attempts to \textit{deregister} by CAS-ing its \emph{location} entry back to \textit{NULL} (\lref{COLLIDE:caslocation1}).
If the CAS is successful, the thread calls the \texttt{activeCollide} function (\lref{COLLIDE:caslocation1success}) in an attempt to either combine or eliminate its operations with those of the other thread. If the CAS fails, however, this indicates that some other thread was quicker and already collided with the current thread; in this case, the current thread becomes a passive thread and executes the \texttt{passiveCollide} function (\lref{COLLIDE:caslocation1fail}).

If the tests of \lref{COLLIDE:checkinfo} fail, the thread attempts to become a passive collider and waits for a short period of time in \lref{COLLIDE:delay} to allow other threads to collide with it. It then tries to deregister by CAS-ing its entry in the \emph{location} array to \emph{NULL}. If the CAS fails - implying that an active collider succeeded in initiating a collision with the delegate thread - the delegate thread, now a passive collider, calls the \texttt{passiveCollide} (\lref{COLLIDE:caslocation2fail}) function in an attempt to finalize the collision. If the CAS succeeds, the thread returns \emph{false} indicating that the operation failed on the elimination-combining layer and should be retried on the central stack.

%%%%%%%%%%%%%%%%%%%%%%%%%%%%

\begin{figure}[t!]
\caption[activeCollide]{(a) The \texttt{activeCollide}, (b) \texttt{passiveCollide} and, (c) \texttt{combine} functions.}
\label{fig:startCollision}

\begin{multicols}{2}

%\NoCaptionOfAlgo
\nocaptionofalgo
\begin{algorithm}[H]
%\SetAlgoLined
\SetLine
%%%%%%%%%%%%%%%%%%%%%%%%%%%%%%%%%%%%%%%%
\caption{(a) \textbf{boolean} \texttt{activeCollide}(multiOp:aInf,pInf)}
%%%%%%%%%%%%%%%%%%%%%%%%%%%%%%%%%%%%%%%%
    \eIf{CAS(\&location$[$pInf.id$]$, pInf, aInf) \nllabel{STARTCOL:cas}}{
	    \eIf{aInf.op = pInf.op \nllabel{STARTCOL:compareOp}}{
    	    \texttt{combine}(\emph{aInf},\emph{pInf})\nllabel{STARTCOL:combine}\;
	        \Return \False\;\nllabel{STARTCOL:returnCombine}
	    }{
	        \texttt{multiEliminate}(\emph{aInf},\emph{pInf})\nllabel{STARTCOL:eliminate}\;
    	    \Return \True\;\nllabel{STARTCOL:returnEliminate}
	    }
    }{
	    \Return \False\;\nllabel{STARTCOL:returnFalse}
    }
\end{algorithm}

\nocaptionofalgo
\begin{algorithm}[H]
%\SetAlgoLined
\SetLine
%%%%%%%%%%%%%%%%%%%%%%%%%%%%%%%%%%%%%%%%
\caption{(c) \texttt{combine}(multiOp: aInf, pInf)}
%%%%%%%%%%%%%%%%%%%%%%%%%%%%%%%%%%%%%%%%
    \If{aInf.op = PUSH \nllabel{COMBINE:compareOp}}{
         \emph{aInf}.\emph{last}.\emph{cell}.\emph{next} = \emph{pInf}.\emph{cell}\nllabel{COMBINE:concatCells}\;
    }
    \emph{aInf}.\emph{last}.\emph{next} = \emph{pInf}\nllabel{COMBINE:concatTinfos}\;
    \emph{aInf}.\emph{last} = \emph{pInf}.\emph{last}\nllabel{COMBINE:setLast}\;
    \emph{aInf}.\emph{length} = \emph{aInf}.\emph{length} + \emph{pInf}.\emph{length}\nllabel{COMBINE:setSize}\;
\end{algorithm}

\nocaptionofalgo
\begin{algorithm}[H]
%\SetAlgoLined
\SetLine
%%%%%%%%%%%%%%%%%%%%%%%%%%%%%%%%%%%%%%%%
\caption{(b) \textbf{boolean} \texttt{passiveCollide}(multiOp:pInf)}
%%%%%%%%%%%%%%%%%%%%%%%%%%%%%%%%%%%%%%%%
    \emph{aInf} = \emph{location}$[$\emph{pInf.id}$]$\nllabel{FINISHCOL:getInfo}\;
    \emph{location}$[$\emph{pInf.id}$]$ = \Null\;\nllabel{FINISHCOL:setInfo}
    \eIf{pInf.op $\neq$ aInf.op \nllabel{FINISHCOL:compareOp}}{
        \If{pInf.op = POP \nllabel{FINISHCOL:isPop}}{
            \emph{pInf}.\emph{cell} = \emph{aInf}.\emph{cell}\nllabel{FINISHCOL:setCell}\;
        }
        \Return \True\;\nllabel{FINISHCOL:returnEliminate}
    }{
        await(\emph{pInf}.\emph{cStatus} $\neq$ INIT)\;\nllabel{FINISHCOL:spin}
		\eIf {pInf.cStatus = FINISHED \nllabel{FINISHCOL:checkFinished}}{
            \Return \True\;\nllabel{FINISHCOL:returnFinished}
        }
        {
            \nllabel{FINISHCOL:retry}
            \emph{pInf}.\emph{cStatus} = INIT\nllabel{FINISHCOL:setInit}\;
            \Return \False\;\nllabel{FINISHCOL:returnFalse}
        }
    }
\end{algorithm}

\end{multicols}

\end{figure}

The \texttt{activeCollide} function (figure \ref{fig:startCollision}-(a)) is called by an active collider in order to attempt to combine or eliminate its operations with those of a passive collider. It receives as its input pointers to the \textit{multiOp} structures of both threads. The active collider first attempts to swap the passive collider's \textit{multiOp} pointer with a pointer to its own \textit{multiOp} structure by performing a \textit{CAS} on the \emph{location} array in \lref{STARTCOL:cas}. If the CAS fails then the passive collider is no longer eligible for collision and the function returns \emph{false} (\lref{STARTCOL:returnFalse}), indicating that the executing thread must retry its multi-op on the central stack. If the \emph{CAS} succeeds, then the collision took place. The active collider now compares the type of its multi-op with that of the passive collider (\lref{STARTCOL:compareOp}) and calls either the \texttt{combine} or the \texttt{multiEliminate} function, depending on whether the multi-ops have identical or reverse semantics, respectively (\llref{STARTCOL:compareOp}{STARTCOL:returnEliminate}). Observe that \texttt{activeCollide} returns \emph{true} in case of elimination and \emph{false} in case of combining. The reason is the following: in the first case it is guaranteed that the executing thread's operation was matched with a reverse-semantics operation and so was completed, whereas in the latter case the operations of the passive collider are delegated to the active collider which must now access the central stack again.

The \texttt{passiveCollide} function (figure \ref{fig:startCollision}-(b)) is called by a passive collider after
it identifies that it was collided with. The passive collider first reads the multi-op pointer written to its entry in the \textit{location} array by the active collider and initializes its entry in preparation for future operations (\llref{FINISHCOL:getInfo}{FINISHCOL:setInfo}). If the multi-ops of the colliding threads-pair are of reverse semantics (\lref{FINISHCOL:compareOp}) then the function returns \emph{true} in \lref{FINISHCOL:returnEliminate}\ because, in this case, it is guaranteed that the colliding delegate threads exchange values. Specifically, if the passive thread's multi-op type is \textit{pop}, the thread copies the cell communicated to it by the active
collider (\lref{FINISHCOL:setCell}).

If both multi-ops are of identical semantics, then the passive collider's operations were delegated to the active thread and the executing thread ceases to be a delegate thread. In this case, the thread waits until it is signalled (by writing to the \emph{cStatus} field of its \emph{multiOp} structure) how to proceed. There are two possibilities: (1) \emph{cStatus = FINISHED} holds in \lref{FINISHCOL:checkFinished}. In this case, the thread's operation response is ready and it returns \emph{true} in \lref{FINISHCOL:returnFinished}. (2) \emph{cStatus = RETRY} holds (\lref{FINISHCOL:retry}) indicating that the executing thread became a delegate thread once again. This occurs if a thread to which the current thread's operation was delegated eliminated with a multi-op that had a shorter list than its own and the first operation in the ``residue'' is the current thread's operation. In this case, the thread changes the value of its $cStatus$ back to \emph{INIT} (\lref{FINISHCOL:setInit}) and returns \emph{false}, indicating that the operation should be retried on the central stack.

The \emph{combine} function (figure \ref{fig:startCollision}-(c)) is called by an active collider when the operations of both colliders have identical semantics. It receives as its input pointers to the \emph{multiOp} structures of the two colliders. It delegates the operations of the passive collider to the active one by concatenating the multiOp list of the passive collider to that of the active collider, and by updating the \emph{last} and \emph{length} fields of its \emph{multiOp} record accordingly (\llref{COMBINE:concatTinfos}{COMBINE:setSize}). In addition, if the type of both multi-ops is \emph{push}, then their cell-lists are also concatenated (\lref{COMBINE:concatCells}); this allows the delegate thread to push all its operations to the central stack by using a single \emph{CAS} operation.

The \emph{multiEliminate} function is called by an active collider when the operations of both colliders have reverse semantics. It matches as many pairs of reverse-semantics operations as possible. If there is a residue of \emph{push} or \emph{pop} operations, it signals the first waiting thread in the residue list by writing the value \emph{RETRY} to the \emph{cStatus} field of its \emph{multiOp} structure. The signaled thread becomes a delegate thread again and retries its multi-op on the central stack. The full pseudo-code of the \emph{multiEliminate} function and its description are deferred to the appendix.

\remove{
\begin{figure}[t!]
\caption[updateInfo]{The updateInfo operation}
\label{fig:updateInfo}

%\NoCaptionOfAlgo
\nocaptionofalgo
\begin{algorithm}[H]
%\SetAlgoLined
\SetLine
%%%%%%%%%%%%%%%%%%%%%%%%%%%%%%%%%%%%%%%%
\caption{updateInfo(multiOp retryInfo, multiOp finishedInfo)}
%%%%%%%%%%%%%%%%%%%%%%%%%%%%%%%%%%%%%%%%
    retryInfo.length=finishedInfo.length\;
    retryInfo.last=finishedInfo.last\;
\end{algorithm}
\end{figure}

}

\section{DECS Performance Evaluation}
\label{sec:evaluation}

\newcommand{\adi}[1]{\textbf{#1}}

We conducted our performance evaluation on a Sun SPARC T5240 machine, comprising two UltraSPARC T2 plus (Niagara II) chips, running the Solaris 10 operating system. Each chip contains 8 cores and each core multiplexes 8 hardware threads, for a total of 64 hardware threads per chip. We ran our experiments on a single chip to avoid communication via the L2 cache. The algorithms we evaluated are implemented in C++ and the code was compiled using GCC with the -O3 flag for all algorithms.

We compare DECS with the Treiber stack\footnote{We evaluated two variants of the Treiber algorithm - with and without exponential backoff. The variant using exponential backoff performed consistently better and is the version we compare with.} and with the most effective known stack implementations: the HSY elimination-backoff stack, and a flat-combining based stack.\footnote{We downloaded the most updated flat-combining code from https://github.com/mit-carbon/Flat-Combining.} \footnote{The Treiber, HSY and DECS algorithms need to cope with the "ABA problem" \cite{IBM370-manual}, since they use dynamic-memory structures that may need to be recycled and perform CAS operations on pointers to these structures. We implemented the simplest and most common ABA-prevention technique that includes a tag with the target memory locations so that both the memory location and the tag are manipulated together atomically, and the tag is incremented with each update of the target memory location \cite{IBM370-manual}.}

In our experiments, threads repeatedly apply operations to the stack for a fixed duration of one second and we measure the resulting \emph{throughput} - the total number of operations applied to the stack - varying the level of concurrency from 1 to 128. Each data point is the average of three runs. We measure throughput on both symmetric (push and pop operations are equally likely) and asymmetric workloads. Stacks are pre-populated with enough cells so that pop operations do not operate on an empty stack also in asymmetric workloads.

%Each algorithm, except for Treiber, uses some technique to eliminate a number of operations (of different threads) %with different semantics by a single thread operation.
%This technique makes these algorithms have an absolute higher throughput value when the workload has a mixed type of %operations, and the highest value is when the number of push and pop operations is equal.
%\adi{Can we support this maybe by showing cache-misses or do you think of any way to support this?}

\begin{figure}[H]
\begin{center}
\captionsetup[subfloat]{labelformat=empty}
%\captionsetup[aboveskip]{0}
\captionsetup[subfloat]{captionskip=-18pt}
%\captionsetup[farskip]{0}
%\captionsetup[position]{top}
    \vspace{-30pt}
    \subfloat[(a) Throughput: 50\% push, 50\% pop]{
        \includegraphics[scale=0.45]{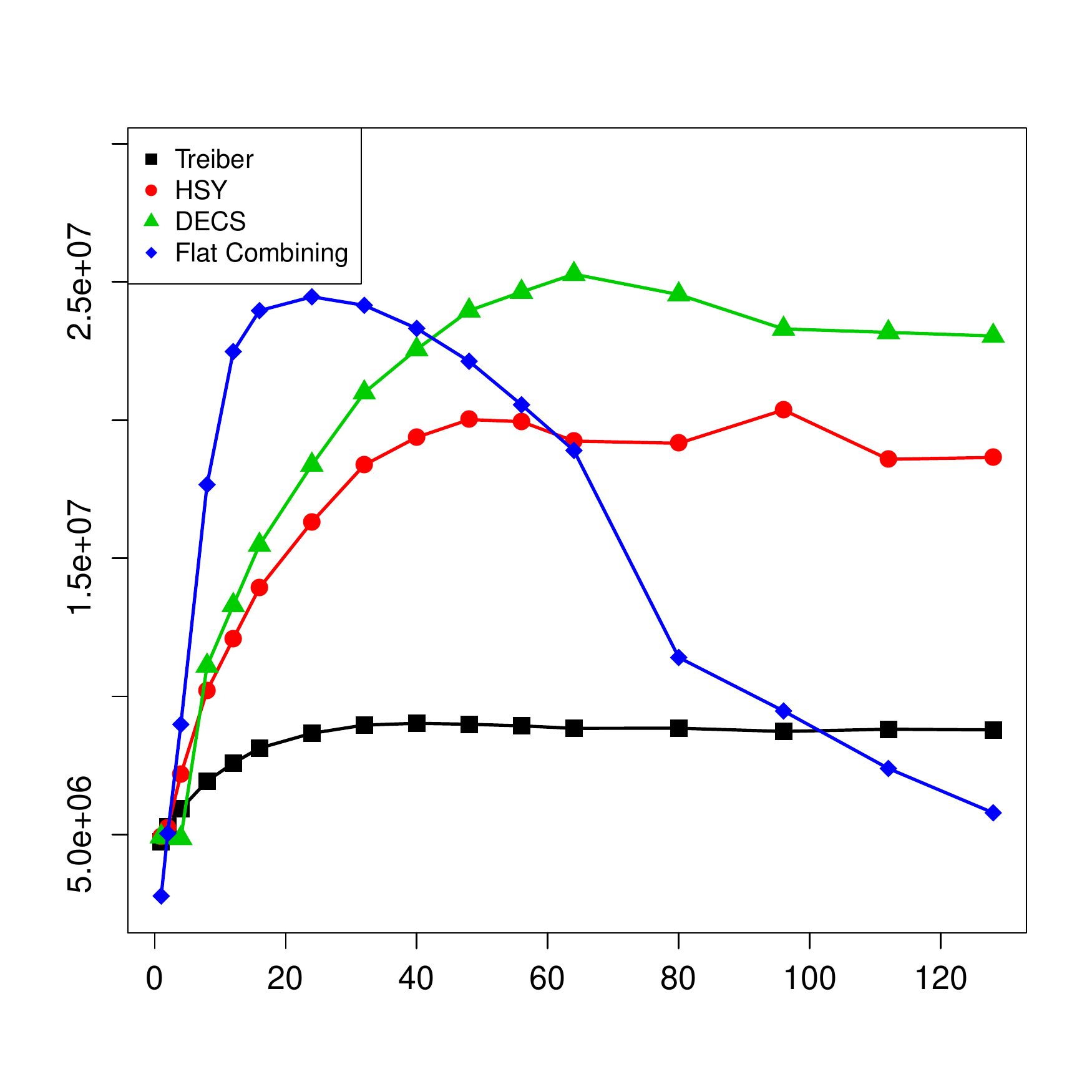}
    } \hspace{-30pt}
    \subfloat[(d) Collision success (\%): 50\% push, 50\% pop]{
        \includegraphics[scale=0.45]{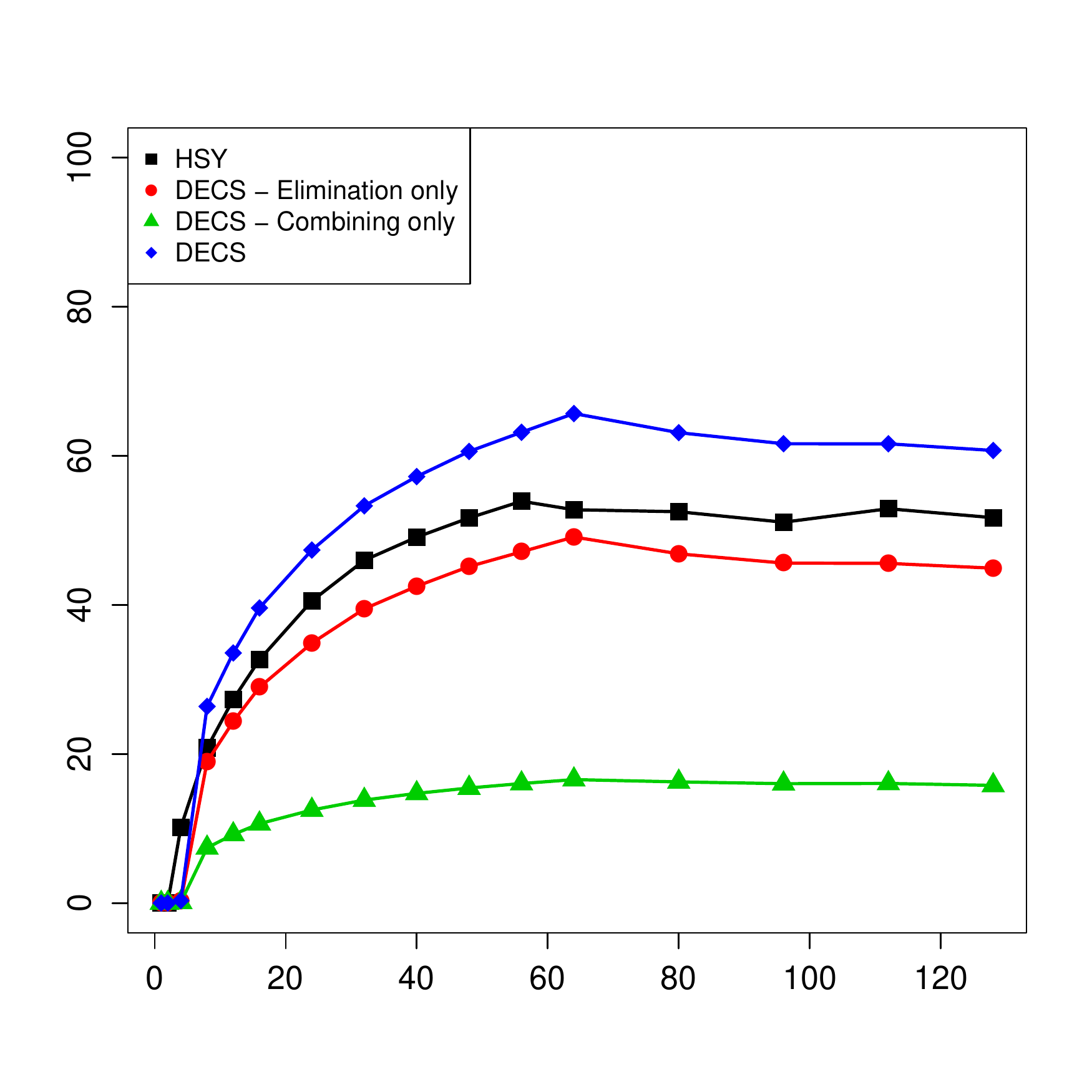}
    } \\
    \vspace{-30pt}
    \subfloat[(b) Throughput: 25\% push, 75\% pop]{
        \includegraphics[scale=0.45]{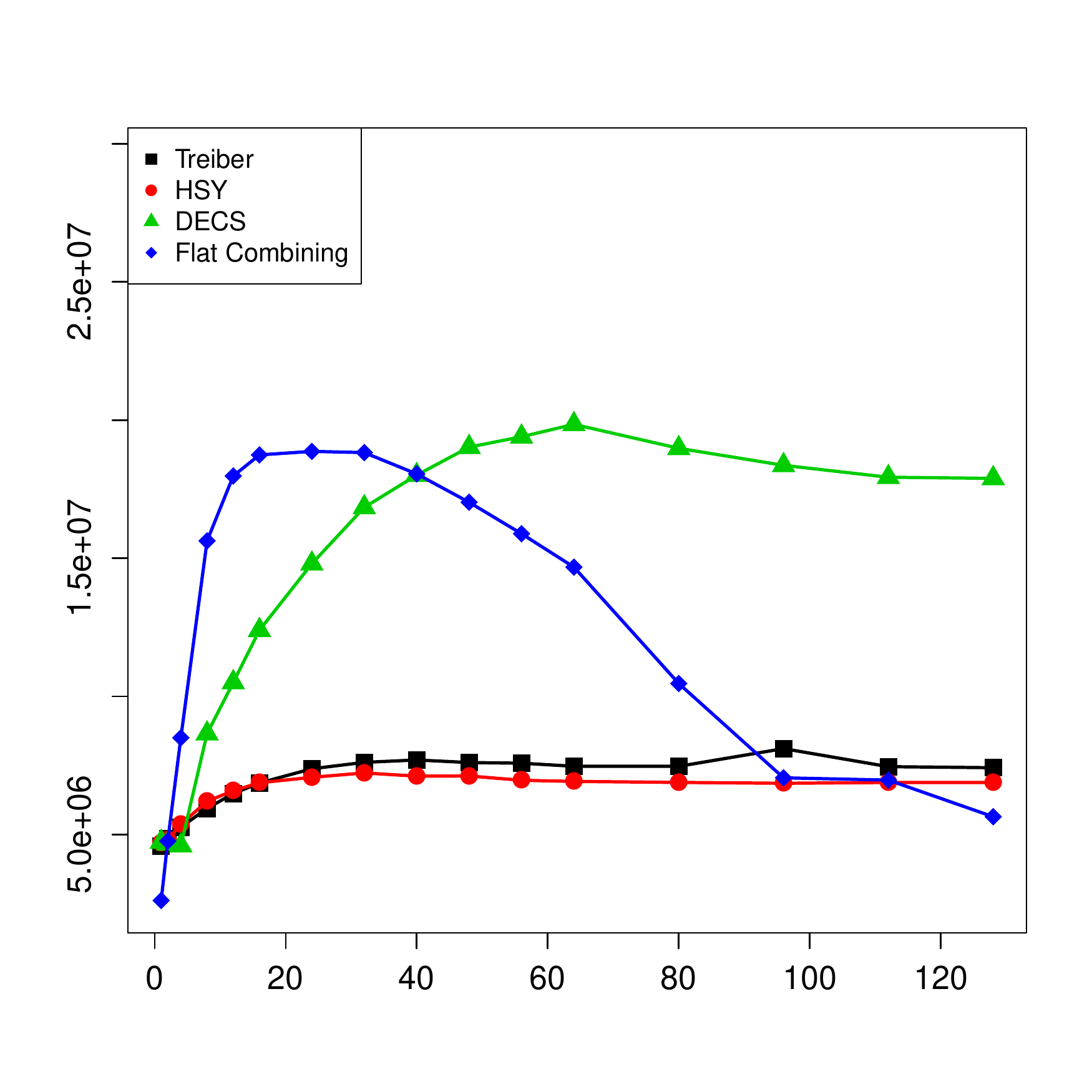}
    } \hspace{-30pt}
    \subfloat[(e) Collision success (\%): 25\% push, 75\% pop]{
        \includegraphics[scale=0.45]{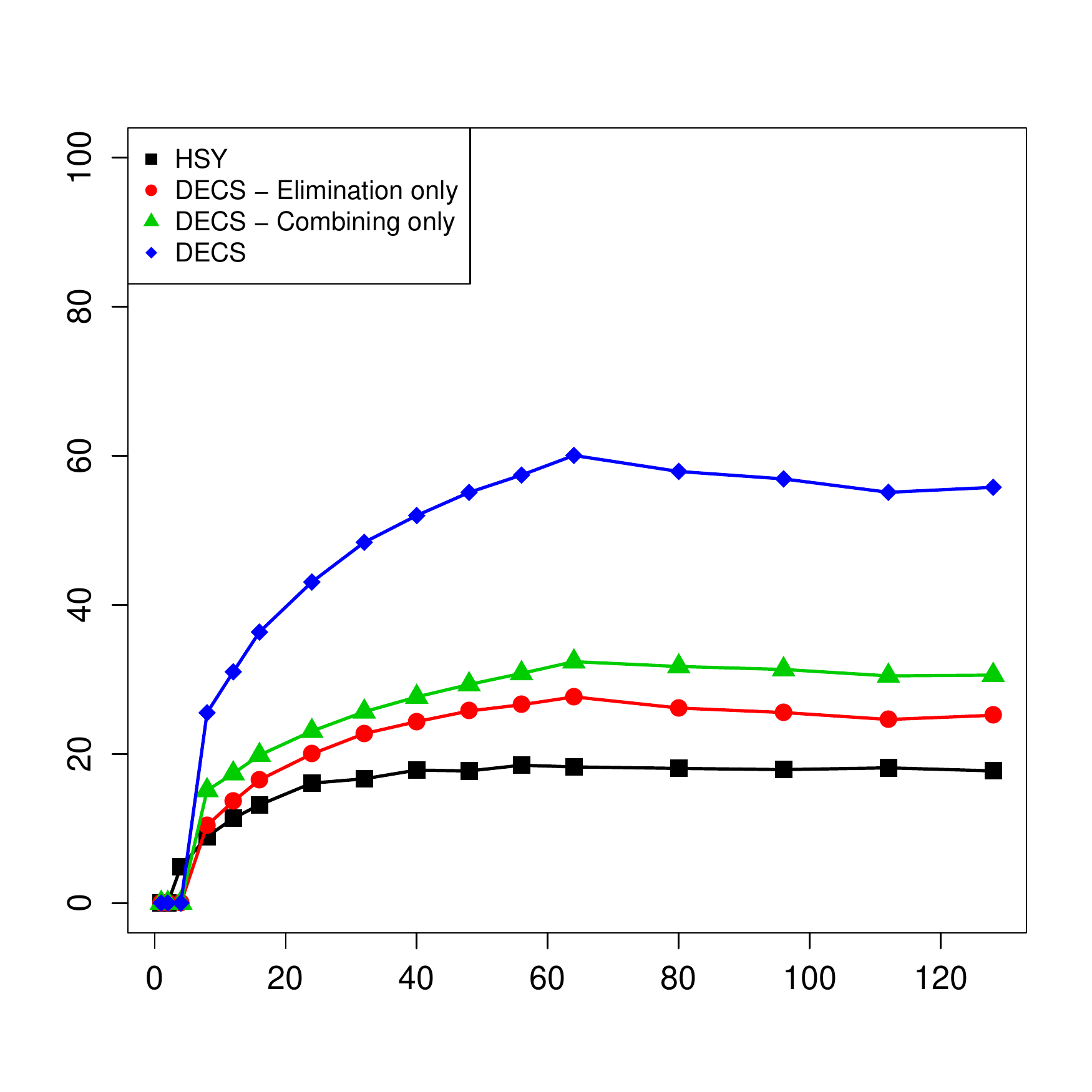}
    } \\
    \vspace{-30pt}
    \subfloat[(c) Throughput: 0\% push, 100\% pop]{
        \includegraphics[scale=0.45]{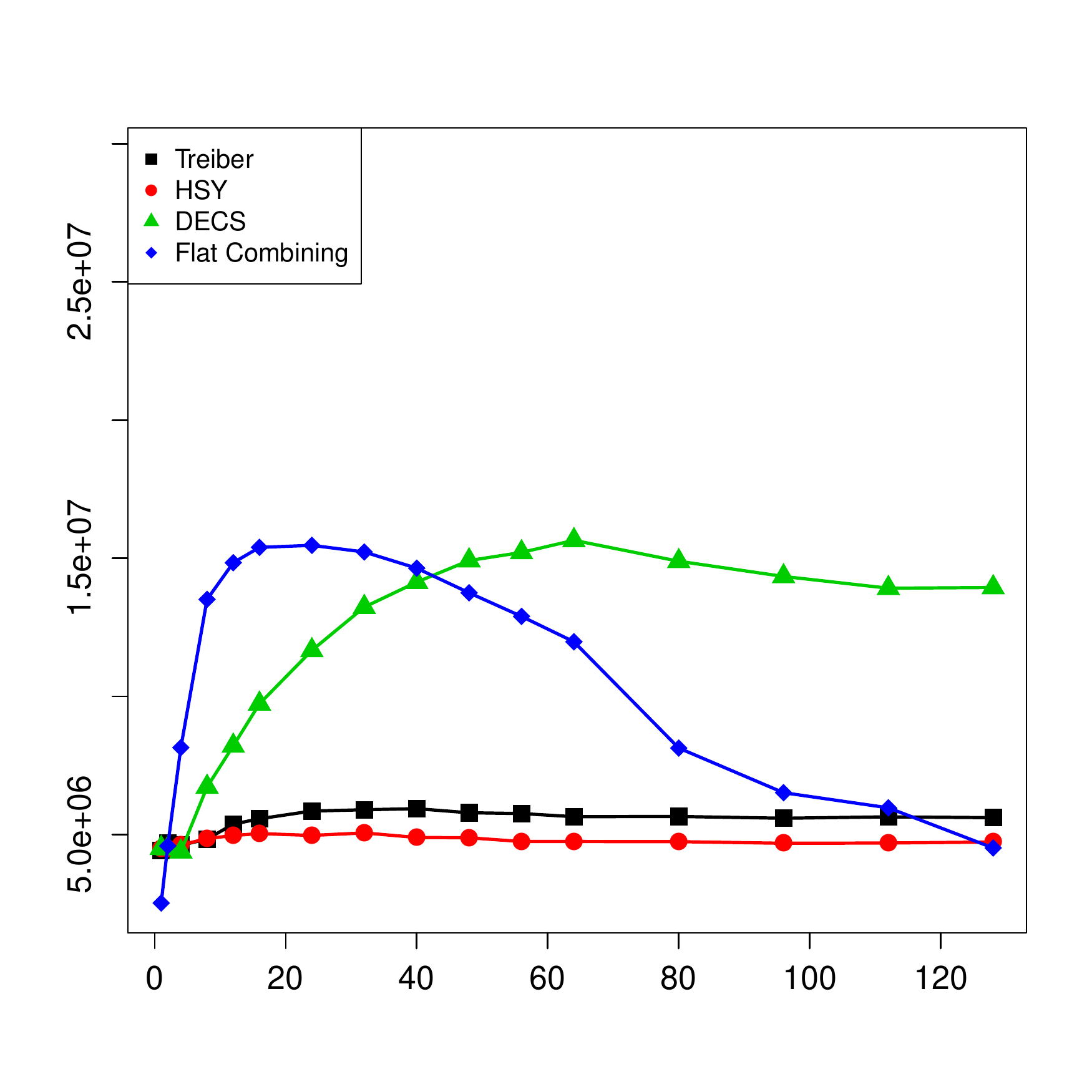}
    } \hspace{-30pt}
    \subfloat[(f) Collision success (\%): 0\% push, 100\% pop]{
        \includegraphics[scale=0.45]{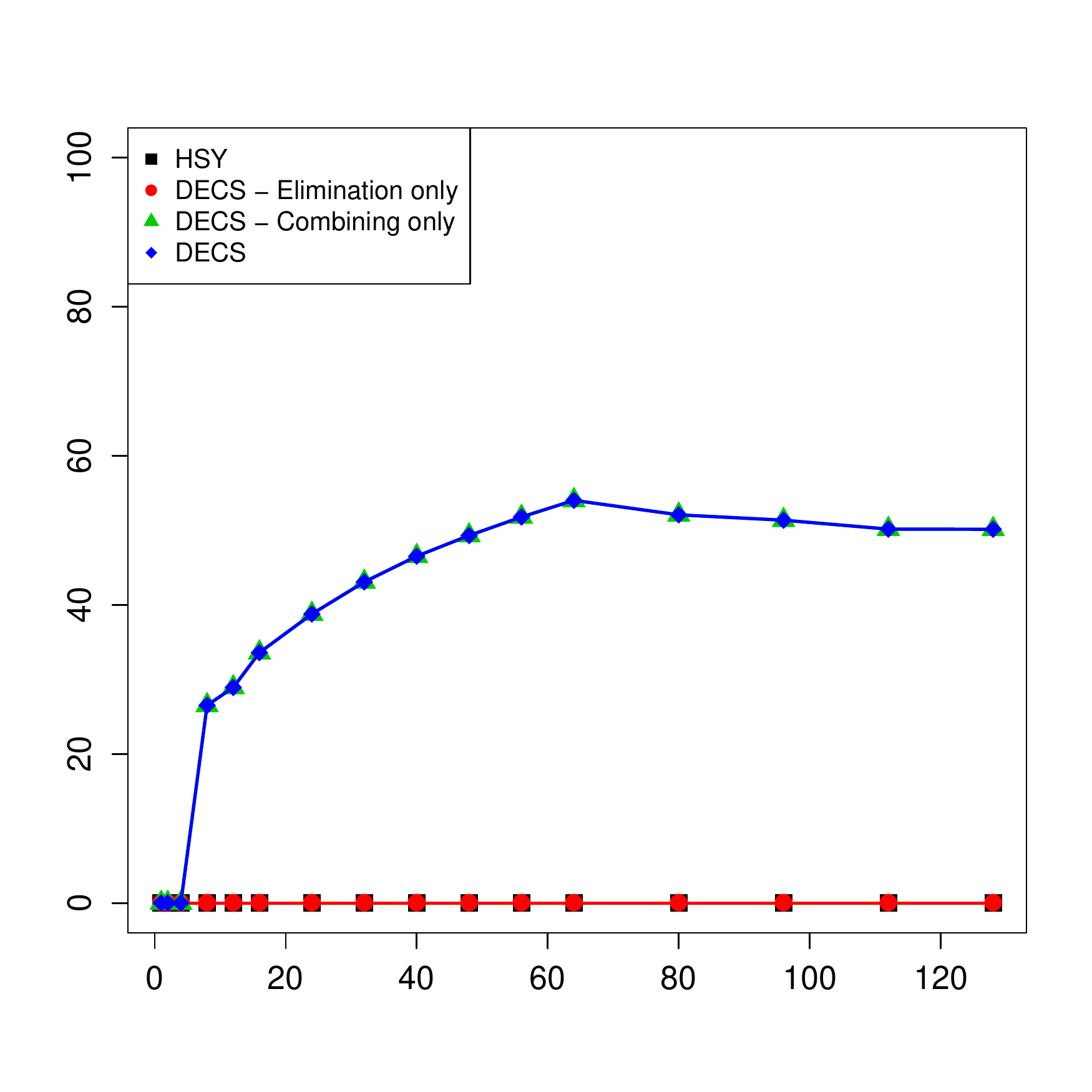}
    }
\captionsetup[subfloat]{labelformat=parens}
\captionsetup[subfloat]{captionskip=4pt}
\end{center}
\caption{Throughput and collision success rates. X-axis: threads \#; Y-axis in (a)-(c): throughput.}
\label{fig:DECSComparisonThroughputCollisionRates}
\end{figure}

Figures \ref{fig:DECSComparisonThroughputCollisionRates}-(a) through (c) compare the throughput of the algorithms we evaluate in symmetric (50\% push, 50\% pop), moderately-asymmetric (25\%push, 75\% pop) and fully-asymmetric (0\% push, 100\% pop) workloads, respectively.
%(For lack of space, larger versions of all graphs and the graphs for the 75\% push and 100\% push workloads are shown in the appendix.)
It can be seen that the DECS stack outperforms both the Treiber stack and the HSY stack for all workload types and all concurrency levels.

\vspace{0.2cm}
\noindent \textbf{Symmetric workloads}
\vspace{0.1cm} \newline
\noindent We first analyze performance on a symmetric workload, which is the optimal workload for the HSY stack. As shown in Figure \ref{fig:DECSComparisonThroughputCollisionRates}-(a), even here the HSY stack is outperformed by DECS by a margin of up to 31\% (when the number of HW threads is 64). This is because, even in symmetric workloads, there is a non-negligible fraction of collisions between operations of identical semantics from which DECS benefits but the HSY stack does not. Both DECS and the HSY stack scale up until concurrency level 64 - the number of hardware threads. When the number of software threads exceeds the number of hardware threads, the HSY stack more-or-less maintains its throughput whereas DECS slightly declines but remains significantly superior to the HSY stack.

The FC stack incurs the highest overhead in the lack of contention (concurrency level 1) because the single running thread still needs to capture the FC lock. Due to its low synchronization overhead it then exhibits a steep increase in its throughput and reaches its peak throughput at 24 threads, where it outperforms DECS by approximately 33\%. The FC stack does not continue to scale beyond this point, however, and
its throughput rapidly deteriorates as the level of concurrency rises. For concurrency levels higher than 40, its performance falls below that of DECS and it is increasingly outperformed by DECS as the level of concurrency is increased: for 64 threads, DECS provides roughly 33\% higher throughput, and for 128 threads DECS outperforms FC by a factor of 4. For concurrency levels higher than 96, the throughput of the FC stack is even lower than that of the Treiber algorithm. The reason for this performance deterioration is clear: the FC algorithm is essentially sequential, since a single thread performs the combined work of other threads. The Treiber algorithm exhibits the worst performance since it is sequential and incurs significant synchronization overhead. It scales moderately until concurrency level 16 and then more-or-less maintains its throughput.

Figure \ref{fig:DECSComparisonThroughputCollisionRates}-(d) provides more insights into the behavior of the DECS and HSY stacks in symmetric workloads. The HSY curve shows the percentage of operations completed by elimination.
The DECS curve shows the percentage of operations not applied directly to the central stack.
\noindent These are the operations completed by either elimination or combining.\footnote{Whenever a muli-op is applied to the central stack, the operation of the delegate thread is regarded as applied directly to the central stack and those of the waiting threads are counted as completed by combining. Similarly, when two multi-ops of reverse semantics collide, the operations of the delegate threads are counted as completed by elimination and those of the waiting threads as completed by combining.}
The curves titled ``Elimination only'' and ``Combining only'' show a finer partition of the DECS operations according to whether they completed through elimination or combining. It can be seen that the overall percentage of operations not completed on the central stack is higher for DECS than for the HSY stack by up to 30\% (for 64 threads), thus reducing the load on the central stack and allowing DECS to perform better than the HSY stack.

\vspace{0.2cm}
\noindent \textbf{Asymmetric workloads}
\vspace{0.1cm} \newline
\noindent Figures \ref{fig:DECSComparisonThroughputCollisionRates}-(b) and \ref{fig:DECSComparisonThroughputCollisionRates}-(c) compare throughput on moderately- and fully-asymmetric workloads, respectively. The relative performance of DECS, the FC and the Treiber stacks is roughly the same as for the symmetric workload; nevertheless, DECS performance decreases because, as can be seen in Figures \ref{fig:DECSComparisonThroughputCollisionRates}-(e) and \ref{fig:DECSComparisonThroughputCollisionRates}-(f), the ratio of DECS operations that complete via elimination is significantly reduced for the 25\% push workload and is 0 for the 0\% push workload. This reduction in elimination is mostly compensated by a corresponding increase in the ratio of DECS operations that complete by combining.

The performance of the HSY stack, however, deteriorates for asymmetric workloads because, unlike DECS, it cannot benefit from collisions between operations with identical semantics. When the workload is moderately asymmetric (Figure \ref{fig:DECSComparisonThroughputCollisionRates}-(b)), the HSY stack scales up to 32 threads but then its performance deteriorates and falls even below that of the Treiber algorithm for 48 threads or more. In these levels of concurrency, the low percentage of successful collisions makes the elimination layer counter-effective. The throughput of the DECS algorithm exceeds that of the HSY stack by a factor of up to 3. The picture is even worse for the HSY algorithm for fully asymmetric workloads (Figure \ref{fig:DECSComparisonThroughputCollisionRates}-(c)), where it performs almost consistently worse than the Treiber algorithm. In these workloads, DECS' throughput exceeds that of the HSY algorithm significantly in all concurrency levels 8 or higher; the performance gap increases with concurrency up until 64 threads and DECS provides about 3 times the throughput for all concurrency levels 64 or higher.

\remove{
When only a single thread is executing operations, all algorithms (except Flat Combining) have the same throughput.
This is because there is no contention, and these algorithms follow Treiber's approach which executes a CAS operation on the central stack and always succeeds.
FC's throughput is lower in the single thread case because of the use of a lock and the overhead of each operation.

Under all workloads Treiber's throughput increases up to 16 threads, and when the threads number is greater than 16 the throughput remains approximately the same.
This is due to a high contention when executing a CAS operation on the stack's top when the number of threads is beyond 16.
In our experiments we also measured the percentage of failed CAS operations (out of total CAS operations).
The results show more than 50\% CAS failure rate for the Treiber algorithm when the threads number is over 16 for all workloads.

One can clearly observe that DECS is at least as good as Treiber and Elimination under all workloads and all concurrency levels.
This is because DECS uses their techniques, when applicable, to perform an operation.
When these techniques are inapplicable (i.e., when there is a high contention and the operations have the same semantics), DECS improves the throughput by combining operations and reducing contention.
We also note that for all workloads DECS reaches its peak at 64 threads.
This is because the system can only support up to 128 hardware threads on 16 cores, which limits concurrency.

} %%% END REMOVE

%%% END REMOVE

\section{The Nonblocking DECS Algorithm}
\label{sec:NB-DECS}

For some applications, a nonblocking stack may be preferable to a blocking one because it is more robust in the face of thread failures. The HSY stack is nonblocking - specifically lock-free \cite{waitfree} - and hence guarantees global progress as long as some threads do not fail-stop. In contrast, both the FC and the DECS stacks are blocking. In this section, we provide a high-level description of NB-DECS, a lock-free variant of our DECS algorithm that allows threads that delegated their operations to another thread and have waited for too long to cancel their ``combining contracts'' and retry their operations. A full description of the algorithm is deferred to the appendix. We also present a comparative evaluation of the new algorithm.

Recall that waiting threads await a signal from their delegate thread in the \texttt{passiveCollide} function (\lref{FINISHCOL:spin} in Figure \ref{fig:startCollision}). In the DECS algorithm, a thread awaits until the delegate thread writes to the \emph{cStatus} field of its \textit{multiOp} structure but may wait indefinitely. In NB-DECS, when a thread concludes that it waited ``long enough'' it attempts to \emph{invalidate} its \emph{multiOp} structure. To prevent race conditions, invalidation is done by applying \emph{test-and-set} to a new \emph{invalid} field added to the \emph{multiOp} structure. A delegate thread, on the other hand, must take care not to assign a cell of a valid \emph{push} operation to an invalid multi-op structure of a \emph{pop} operation. This raises the following complications which NB-DECS must handle.

\begin{enumerate}
\item A delegate thread may pop invalid cells from the central stack. Therefore, in order not to assign an invalid cell to a \emph{pop} operation, the delegate thread must apply \textit{test-and-set} to each popped cell to verify that it is still valid (and if so to ensure it remains valid), which hurts performance.

\item A delegate thread performing a pop multi-op must deal with situations in which some of its waiting threads invalidated their multi-op structures. If the delegate thread were to pop from the central stack more cells than can be assigned to valid multi-op structures in its list, linearizability would be violated. Consequently, unlike in DECS, the delegate thread must pop items from the central stack \emph{one by one}, which also hurts the performance of NB-DECS as compared with DECS.

\item The \texttt{multiEliminate} function, called by an active delegate thread when it collides with a thread with reverse semantics, must also verify that valid cells are only assigned to valid pop multi-ops. Once again, \emph{test-and-set} is used to prevent race conditions.

\end{enumerate}

\begin{figure}[H]
\begin{center}
\captionsetup[subfloat]{labelformat=empty}
\captionsetup[subfloat]{captionskip=-18pt}
	\vspace{-30pt}
	\subfloat[(a) Throughput: 50\% push, 50\% pop]{
		\includegraphics[scale=0.45]{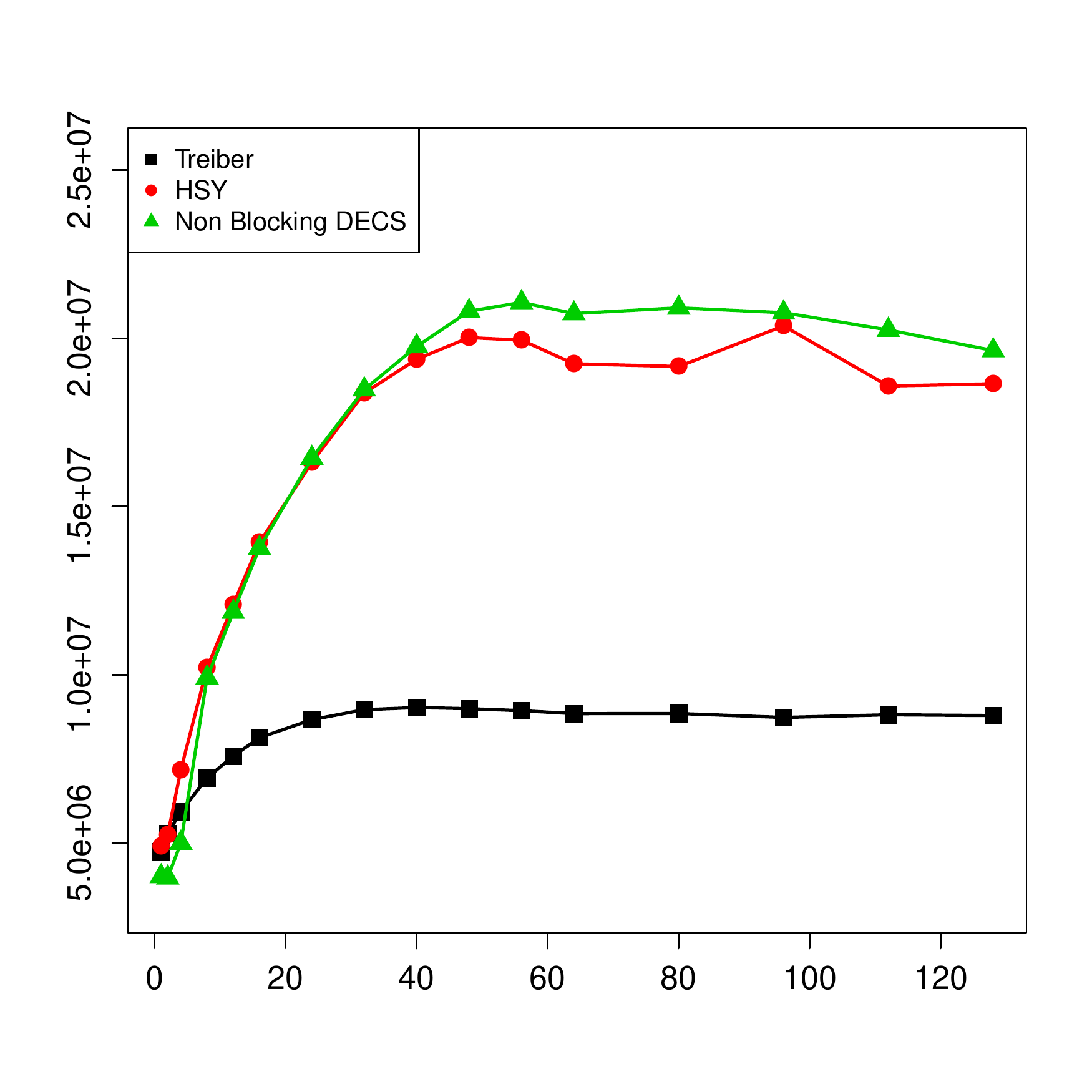}
	} \hspace{-30pt}
	\subfloat[(c) Throughput: 25\% push, 75\% pop]{
		\includegraphics[scale=0.45]{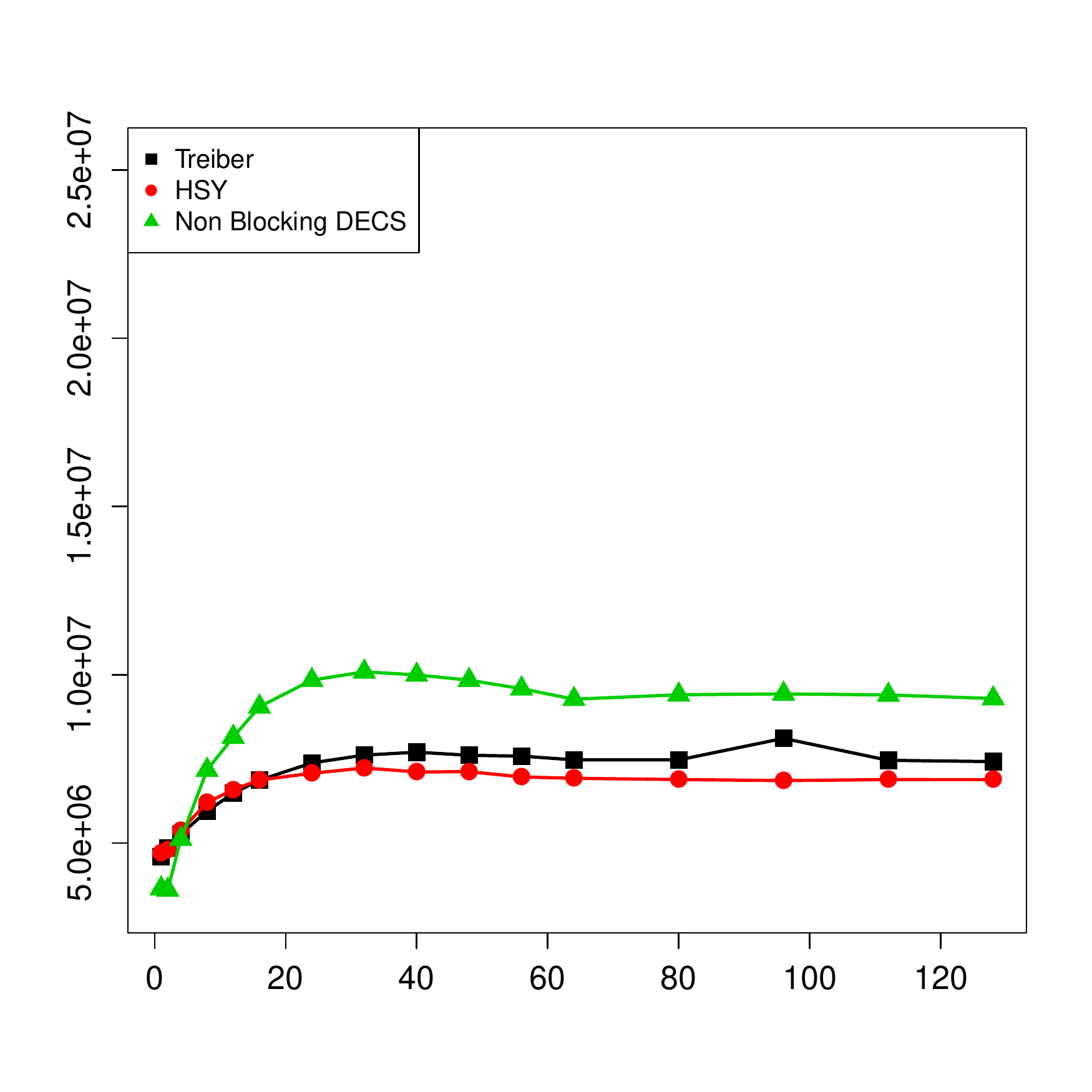}
	} \\
	\vspace{-30pt}
	\subfloat[(b) Throughput: 75\% push, 25\% pop]{
		\includegraphics[scale=0.45]{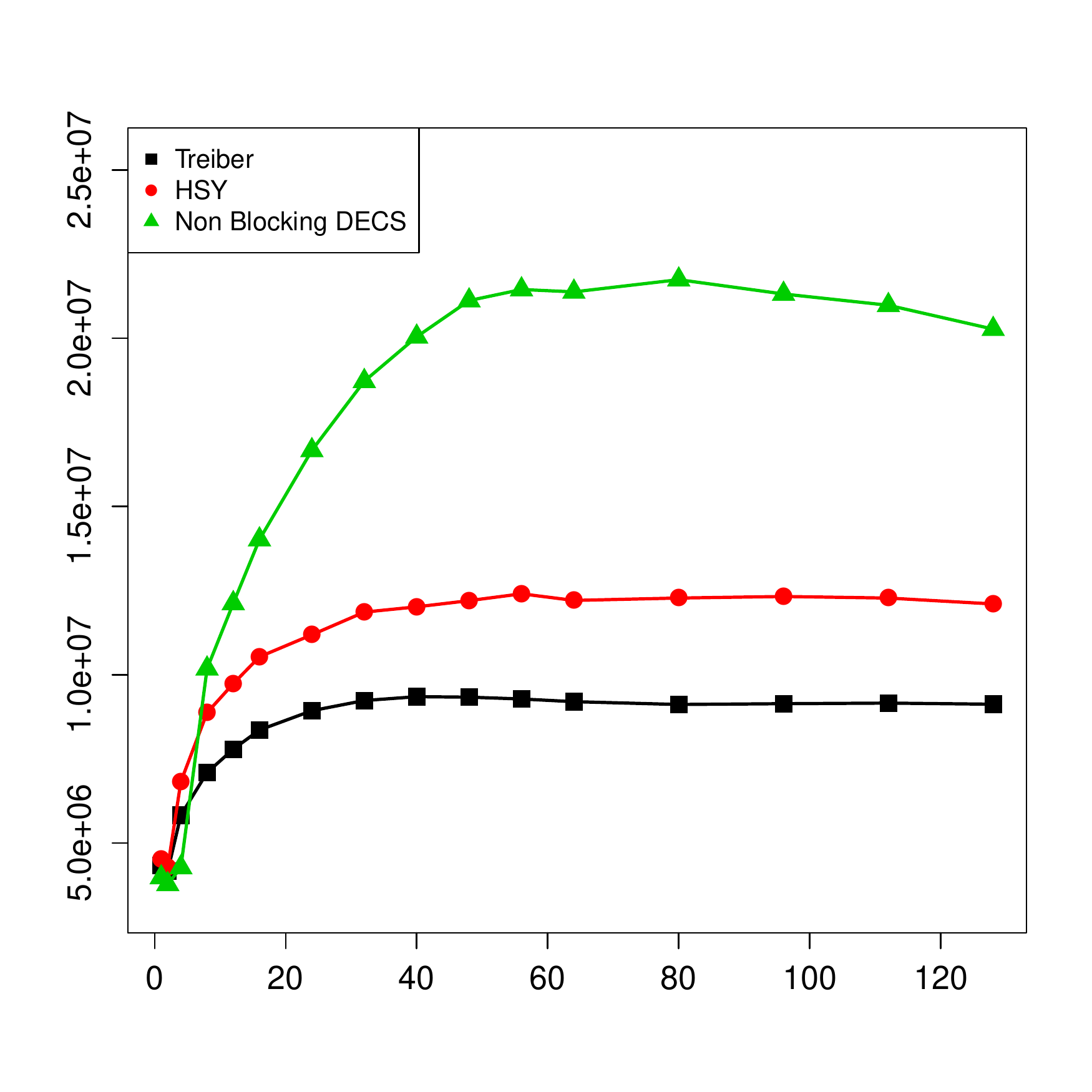}
	} \hspace{-30pt}
	\subfloat[(d) Throughput: 100\% push, 0\% pop]{
		\includegraphics[scale=0.45]{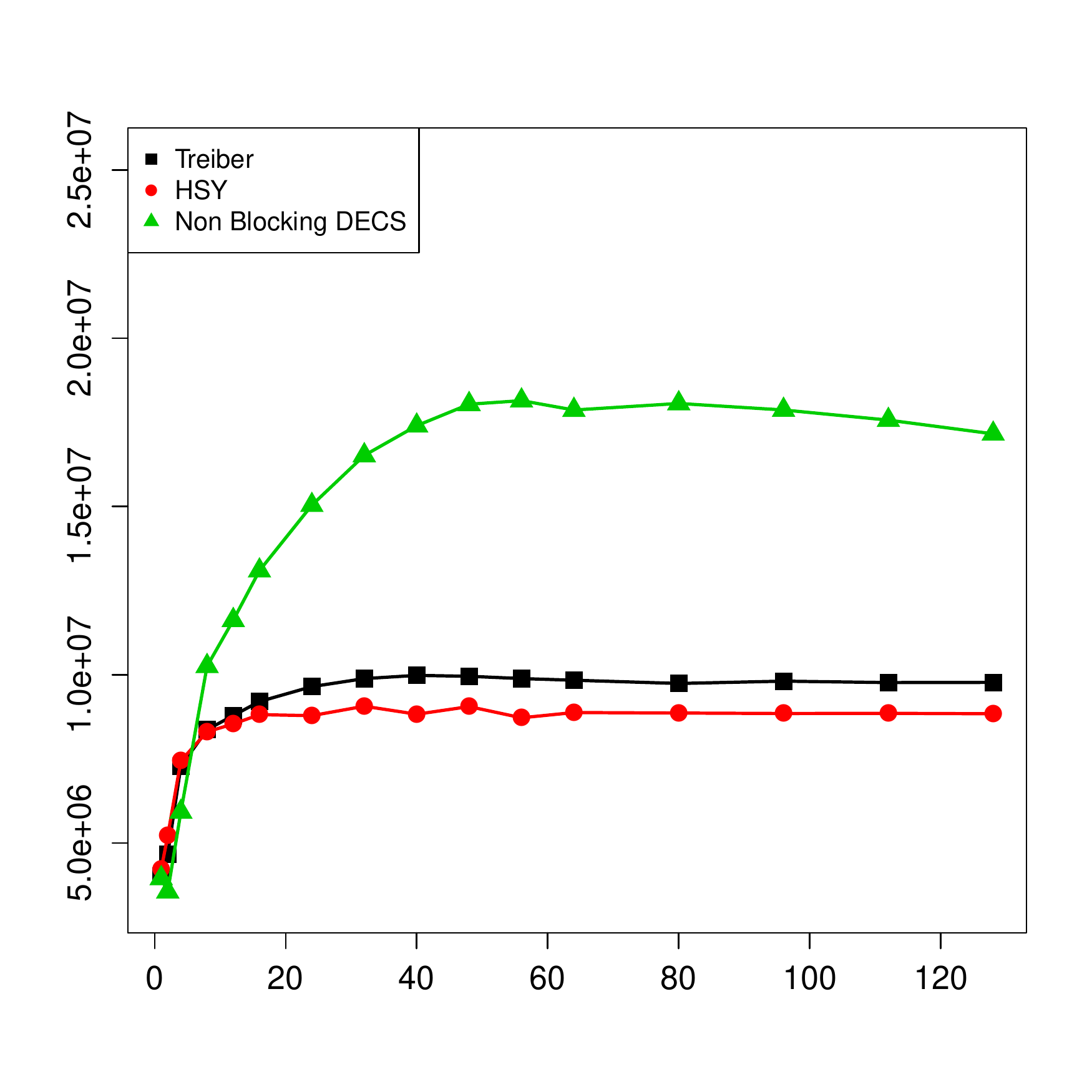}
	}
\captionsetup[subfloat]{labelformat=parens}
\captionsetup[subfloat]{captionskip=4pt}
\end{center}
\caption{Throughput of lock-free algorithms. X-axis: threads \#; Y-axis: throughput.}
\label{fig:NB-DECSComparisonThroughput}
\end{figure}

\vspace{0.2cm}
\noindent \textbf{NB-DECS performance evaluation.}
\vspace{0.1cm} \newline
\noindent Due to the extra synchronization introduced in NB-DECS for allowing threads to invalidate operations that are pending for too long, the throughput of NB-DECS is, in general, significantly lower than that of the (blocking) DECS stack. We compare NB-DECS with two other lock-free algorithms: Treiber and the HSY stack. As shown in Figure \ref{fig:NB-DECSComparisonThroughput}, the performance of the NB-DECS and HSY stacks on symmetric workloads is almost identical (Figure \ref{fig:NB-DECSComparisonThroughput}-(a)), with a slight advantage to NB-DECS for concurrency levels of 36 or more, and they both scale significantly better than the Treiber stack.

For moderately-asymmetric workloads, NB-DECS performs much better than the HSY stack but its advantage is much greater when there is a majority of \emph{push} operations. The reason for this difference is that the extra synchronization added to NB-DECS (as compared with DECS) hurts \emph{pop} operations more than it does \emph{push} operations. Specifically, complication 2. above hurts the performance of multi-pop operations applied to the central stack (since they need to pop cells one by one) but multi-push operations to the central stack may still push their entire list atomically. For workloads with 75\% \emph{push} operations (Figure \ref{fig:NB-DECSComparisonThroughput}-(b)), NB-DECS outperforms the HSY stack significantly and the margin increases with concurrency. Specifically, for 56 threads or more, NB-DECS outperforms HSY by more than 70\%. For workloads with 25\% \emph{push} (Figure \ref{fig:NB-DECSComparisonThroughput}-(b)), the difference is smaller but still significant and NB-DECS outperforms HSY by about 35\% for concurrency levels 24 or higher.

The state of affairs is similar for fully-asymmetric workloads. When the workload consists of \emph{push} operations only (Figure \ref{fig:NB-DECSComparisonThroughput}-(d)), NB-DECS scales nicely up to 56 threads and then more-or-less retains its throughput, delivering performance of up to 2 times that of the HSY stack. On the other hand, when the workload consists of only \emph{pop} operations
%(see Figure \ref{fig:appendix-NB-DECS-througput} in the appendix) 
the extra synchronization hurts NB-DECS and the difference in performance is much smaller.

\section{Discussion}
\label{sec:discussion}

We present DECS, a novel Dynamic Elimination-Combining Stack algorithm. Our empirical evaluation shows
that DECS scales significantly better than (both blocking and nonblocking) best known stack algorithms for all workload types, providing throughput that is significantly superior to that of both the elimination-backoff stack and the flat-combining stack for high concurrency levels. We also present NB-DECS - a lock-free variant of DECS. NB-DECS provides lower throughput than (the blocking) DECS due to the extra synchronization required for satisfying lock-freedom but may be preferable for some applications since it is more robust to thread failures. NB-DECS significantly outperforms the elimination-backoff stack, the most scalable prior lock-free stack on almost all workload types. The key feature that makes DECS highly effective is the use of a dynamic elimination-combining layer as a backoff scheme for a central data-structure. We believe that this idea may be useful for obtaining high-performance implementations of additional concurrent data-structures. \\

\noindent \emph{Acknowledgements:} We thank Yehuda Afek and Nir Shavit for allowing us to use their Sun SPARC T5240 machine. 

\newpage
\bibliographystyle{abbrv}
\bibliography{paper}

\appendix
\newpage
\section*{Appendix A: DECS Pseudo-Code Omitted From Paper Body}

Figure \ref{fig:mainops} presents the code performed by a thread when it applies a \emph{push} or a \emph{pop} operation to the DECS stack. A \emph{pop} (\emph{push}) operation starts by initializing a \emph{multiOp} record in \lref{MAIN:popinit} (\lref{MAIN:pushinit}). It then attempts to apply the \emph{pop} (\emph{push}) operation to the central stack in \lref{MAIN:popcentral} (\lref{MAIN:pushcentral}).
If this attempt fails, the thread then attempts to apply its operation to the elimination-combining layer in \lref{MAIN:popcollide} (\lref{MAIN:pushcollide}).
A delegate thread continues these attempts repeatedly until it succeeds. A \emph{pop} operation returns the data stored at the cell that it received either from the central stack (\lref{MAIN:popcentralret}) or by way of elimination (\lref{MAIN:popcollideret}).

\begin{figure}[H]
\caption[DECS Functions]{DECS \texttt{Push} and \texttt{Pop} functions.}
\label{fig:mainops}

%\NoCaptionOfAlgo
\nocaptionofalgo
\linesnumbered
\begin{algorithm}[H]
%\SetAlgoLined
\SetLine
%%%%%%%%%%%%%%%%%%%%%%%%%%%%%%%%%%%%%%%%
\caption{\textbf{Data} \texttt{pop}()}
%%%%%%%%%%%%%%%%%%%%%%%%%%%%%%%%%%%%%%%%
	\multiOp mOp = initMultiOp()\nllabel{MAIN:popinit}\;
    \While{\True}{
        \uIf{cMultiPop(mOp) \nllabel{MAIN:popcentral}}{
            \Return \emph{mOp}.\emph{cell}.\emph{data}\nllabel{MAIN:popcentralret}\;
        }\ElseIf{collide(mOp) \nllabel{MAIN:popcollide}}{
            \Return \emph{mOp}.\emph{cell}.\emph{data}\nllabel{MAIN:popcollideret}\;
        }
    }
\end{algorithm}

%\NoCaptionOfAlgo
\nocaptionofalgo
\begin{algorithm}[H]
%\SetAlgoLined
\SetLine
%%%%%%%%%%%%%%%%%%%%%%%%%%%%%%%%%%%%%%%%
\caption{\texttt{push}(\emph{data})}
%%%%%%%%%%%%%%%%%%%%%%%%%%%%%%%%%%%%%%%%
	\multiOp mOp = initMultiOp(\emph{data})\nllabel{MAIN:pushinit}\;
    \While{\True}{
        \uIf{cMultiPush(mOp) \nllabel{MAIN:pushcentral}}{
            \Return\;
        }\ElseIf{collide(mOp) \nllabel{MAIN:pushcollide}}{
            \Return;\nllabel{MAIN:pushcollideret}
        }
    }
\end{algorithm}
\end{figure}

Figure \ref{fig:multiEliminate} presents the pseudo-code of the \emph{multiEliminate} function, which is called by an active collider when the operations of both colliders have reverse semantics. It receives as input pointers to the \emph{multiOp} records of the active and passive colliders. In the loop of \llref{ELIMINATE:repeatCond}{ELIMINATE:endRepeat}, as many pairs of reverse-semantics operations as possible are matched until at least one of the operation lists is exhausted. All matched operations are signalled by writing the value \emph{FINISHED} to the \emph{cStatus} field of their \emph{multiOp} structure, indicating that they can terminate (\llref{ELIMINATE:activeOpFinished}{ELIMINATE:passiveOpFinished}). Note that both lists contain at least one operation, thus at least a single pair of operations are matched. If the lengths of the multi-ops are unequal, then a ``residue'' sublist remains. In this case, the \textit{length} and \emph{last} fields of the \emph{multiOp} structure belonging to the first waiting thread in the residue sub-list are set. Then that thread is signalled by writing the value \emph{RETRY} to the \emph{cStatus} field of its \emph{multiOp} structure (in \lref{ELIMINATE:aCurrRetry} or \lref{ELIMINATE:pCurrRetry}). This makes the signaled thread a delegate thread once again and it will retry its multi-op on the central stack.

\begin{figure}[h]
\caption[combineEliminate]{The \texttt{multiEliminate} function.}
\label{fig:multiEliminate}

\nocaptionofalgo
\begin{algorithm}[H]
%\SetAlgoLined
\SetLine
%%%%%%%%%%%%%%%%%%%%%%%%%%%%%%%%%%%%%%%%
\caption{\texttt{multiEliminate}(multiOp: \emph{aInf}, \emph{pInf})}
%%%%%%%%%%%%%%%%%%%%%%%%%%%%%%%%%%%%%%%%
	\emph{aCurr} = \emph{aInf}\nllabel{ELIMINATE:initACurr}\;
	\emph{pCurr} = \emph{pInf}\nllabel{ELIMINATE:initPCurr}\;
	\Repeat{aCurr = \Null \OrCond pCurr = \Null\nllabel{ELIMINATE:endRepeat}}{\nllabel{ELIMINATE:repeatCond}
		\eIf{aInf.op = POP \nllabel{ELIMINATE:activePop}}{
			\emph{aCurr}.\emph{cell} = \emph{pCurr}.\emph{cell}\nllabel{ELIMINATE:matchCell1}\;
		}{
			\emph{pCurr}.\emph{cell} = \emph{aCurr}.\emph{cell}\nllabel{ELIMINATE:matchCell2}\;
		}\nllabel{ELIMINATE:endIf}
		\emph{aCurr}.\emph{cStatus} = FINISHED\nllabel{ELIMINATE:activeOpFinished}\;
		\emph{pCurr}.\emph{cStatus} = FINISHED\nllabel{ELIMINATE:passiveOpFinished}\;
		\emph{aInf}.\emph{length} = \emph{aInf}.\emph{length} - 1\nllabel{ELIMINATE:reduceActiveLength}\;
		\emph{pInf}.\emph{length} = \emph{pInf}.\emph{length} - 1\nllabel{ELIMINATE:reducePassiveLength}\;
		\emph{aCurr} = \emph{aCurr}.\emph{next}\nllabel{ELIMINATE:nextA}\;
		\emph{pCurr} = \emph{pCurr}.\emph{next}\nllabel{ELIMINATE:nextP}\;
	}
	\uIf{aCurr $\neq$ \Null \nllabel{ELIMINATE:checkACurr}}{
		\emph{aCurr}.\emph{length} = \emph{aInf}.\emph{length}\nllabel{ELIMINATE:aCurrLength}\;
		\emph{aCurr}.\emph{last} = \emph{aInf}.\emph{last}\;
%		updateInfo(aCurr, aInfo)\nllabel{ELIMINATE:updateActive}\;
		\emph{aCurr}.\emph{cStatus} = RETRY\nllabel{ELIMINATE:aCurrRetry}\;
	}\ElseIf{pCurr $\neq$ \Null \nllabel{ELIMINATE:checkPCurr}}{
		\emph{pCurr}.\emph{length} = \emph{pInf}.\emph{length}\nllabel{ELIMINATE:pCurrLength}\;
		\emph{pCurr}.\emph{last} = \emph{pInf}.\emph{last}\;
%		updateInfo(pCurr, pInfo)\nllabel{ELIMINATE:updatePassive}\;
		\emph{pCurr}.\emph{cStatus} = RETRY\nllabel{ELIMINATE:pCurrRetry}\;
	}
\end{algorithm}
\end{figure}

\newpage

\section*{Appendix B: DECS Correctness Proof Sketch}

A concurrent stack is a data structure whose operations are linearizable to those of the sequential stack.

\begin{definition}
\label{def:pool}
A Stack S is an object that supports two operation types: pop and push. The state of the stack is a sequence of items \emph{S} = $<$$v_0$, ....,$v_n$$>$.\\
The stack is initially empty. The pop and push operations induce the following state transitions to S with appropriate return values.
\begin{itemize}
\item \emph{pop()}: if \emph{S} is not empty, returns $v_n$ and changes \emph{S} to $<$$v_0$, ....,$v_{n-1}$$>$, otherwise returns empty and \emph{S} remains unchanged.
\item \emph{push($v_{new}$)}: changes \emph{S} to $<$$v_0$,....,$v_n$, $v_{new}$$>$.
\end{itemize}
\end{definition}

A pool is a relaxation of a stack that does not require LIFO ordering. Similarly to the proofs in \cite{DBLP:journals/jpdc/HendlerSY10}, we start by proving that DECS has correct pool semantics and then prove that it is linearizable to a sequential stack. Finally we prove that DECS is deadlock-free.

\subsection*{Correct pool semantics}

\begin{definition}
\label{def:pool}
A stack algorithm has correct pool semantics if the following requirements are met for all stack operations:
\begin{enumerate}
\item Let $Op$ be a pop operation, then if the number of push operations preceding $Op$ is larger than the number of pop operations preceding it, $Op$ returns an item (a non-empty value).
\item Let $Op$ be a pop operation that returns an item i, then i was previously pushed by a push operation.
\item Let $Op$ be a push operation that pushed an item i to the stack, then there is at most a single pop operation that returns it.
\end{enumerate}
An operation that complies with the above requirements is called a correct pool operation.
\end{definition}

%%%%%%%%%%%%%%%%%%%%%%%%%%%%%%%%%%%%%%%%%%%%%%%%%%%%%%%%%%%%%%%%%%%%%%%%%%%%%%%%%%%%%%%%%%%%%%%%%%%%%%%%%%%%%%%%%%%%%%%%%%%%%%%%%%%%%%%%%%%%%%%%%%%%%%%%%%%%

\begin{definition}
\label{def:collidingOperation1}
We say that a multi-operation is a \emph{colliding multi-op} if it returns in \lref{COLLIDE:caslocation1success}, \lref{COLLIDE:caslocation1fail} or \lref{COLLIDE:caslocation2fail} (after executing \texttt{activeCollide} or \texttt{passiveCollide}).
\end{definition}

\begin{definition}
\label{def:collidingOperation2}
Let $op_1$ and $op_2$ two multi-operations. We say that $op_1$ and $op_2$ have \emph{collided} if one of the following conditions hold:
\begin{itemize}
\item  $op_1$ performs a successful CAS in \lref{STARTCOL:cas} of \texttt{activeCollide} and pInf points to the multiOp of $op_2$ at that time.
\item $op_2$ performs a CAS in \lref{COLLIDE:caslocation2} of \texttt{collide} and the CAS fails because the entry corresponding to $op_2$ in the location array points to the multiOp of $op_1$.
\end{itemize}
\end{definition}

\begin{definition}
\label{def:collidingOperation}
If op is a colliding multi-op that executes a successful CAS in  \lref{STARTCOL:cas} we say it is an \emph{active colliding multi-op}. If it performs an unsuccessful CAS operation in \lref{COLLIDE:caslocation1} or \lref{COLLIDE:caslocation2} we say it is a \emph{passive colliding multi-op}.
\end{definition}

\begin{lemma}
\label{lemma:activeOrPassive}
Every colliding multi-op is either passive or active, but not both.
\end{lemma}
\begin{proofsketch}
First, a multi-op cannot collide with itself since it verifies in \lref{COLLIDE:checkinfo} that the other multi-op has a different id. If $op$ is an active colliding multi-op, then from Definition \ref{def:collidingOperation} it executes a successful CAS in \lref{STARTCOL:cas}. It follows from the code that an active colliding multi-op executes the \texttt{activeCollide} function, implying that it executed a successful CAS in \lref{COLLIDE:caslocation1} and by Definition \ref{def:collidingOperation} that means $op$ is not a passive colliding multi-op.

A similar argument proves that if $op$ is a passive colliding multi-op then it is not an active colliding multi-op.
\end{proofsketch}

\begin{lemma}
\label{lemma:collidesWithOne}
Every passive colliding multi-op collides with exactly one active colliding multi-op and vice versa.
\end{lemma}
\begin{proofsketch}
Let $op_1$ be an active colliding multi-op. According to Lemma \ref{lemma:activeOrPassive}, $op_1$ cannot be a passive colliding multi-op. From Definition \ref{def:collidingOperation}, $op_1$ performed a successful CAS in \lref{STARTCOL:cas}.
Let $op_2$ denote the multiOp that $op_1$ sets in its corresponding location array entry on \lref{STARTCOL:cas}.
Only $op_2$ could have written its multiOp in its corresponding entry in the location array on \lref{COLLIDE:announce}.
Hence, $op_2$ is executing \texttt{collide} and has not yet finished its execution because the CAS on \lref{COLLIDE:caslocation2} was not performed. Since $op_1$ performed a successful CAS in \lref{STARTCOL:cas}, the CAS performed by $op_2$ on \lref{COLLIDE:caslocation1} or on \lref{COLLIDE:caslocation2} fails. It follows from Definition \ref{def:collidingOperation} that $op_2$ is a passive colliding multi-op.
Finally, any other operation trying to perform the CAS operation in \lref{STARTCOL:cas} on the location of the process executing $op_2$ will fail, hence $op_1$ is the only active colliding multi-op that collides with $op_2$.
%adi{Shouldn't this be proved by the way of contradiction?}
\end{proofsketch}

\begin{lemma}
\label{lemma:activeIsWritten}
Let $op_2$ be a passive colliding multi-op and let $op_1$ be the active colliding multi-op it collided with. Then when $op_2$ enters \texttt{passiveCollide}, a pointer to the multiOp of $op_1$ is written in location$[op_2]$.
\end{lemma}
\begin{proofsketch}
$op_2$ enters \texttt{passiveCollide} after an unsuccessful CAS on \lref{COLLIDE:caslocation1} or \lref{COLLIDE:caslocation2}.
This could have happened only if $op_1$ performed a successful CAS on \lref{STARTCOL:cas} and set location$[op_2]$ to \emph{pInf}, pointing to $op_2$'s multiOp and setting location$[op_2]$.
\end{proofsketch}

%%%%%%%%%%%%%%%%%%%%%%%%%%%%%%%%%%%%%%%%%%%%%%%%%%%%%%%%%%%%%%%%%%%%%%%%%%%%%%%%%%%%%%%%%%%%%%%%%%%%%%%%%%%%%%%%%%%%%%%%%%%%%%%%%%%%%%%%%%%%%%%%%%%%%%%%%%%%
%waiting operations and delegate operation.

\begin{definition}
\label{def:collisionType}
let $op_1$ be an active colliding multi-op and $op_2$ be the passive colliding multi-op it collided with.
If both operations have the same operation semantics, we say the collision is a \emph{combining collision} and we call $op_1$ and $op_2$ \emph{combining multi-ops}. Otherwise we say the collision is an \emph{eliminating collision} and we call $op_1$ and $op_2$ \emph{eliminating multi-ops}.
\end{definition}

\begin{definition}
\label{def:waitingOperations}
We say that a multi-op is a \emph{waiting} multi-op if it executes \lref{FINISHCOL:spin}.
\end{definition}

\begin{lemma}
\label{lemma:waitingIsCombining}
A waiting multi-op is a passive colliding multi-op that collided with an active multi-op
of the same semantics.
\end{lemma}
\begin{proofsketch}
From the \texttt{collide} function and Definition \ref{def:collidingOperation} it follows that an operation executes \texttt{passiveCollide} iff it is a passive colliding multi-op.
From Lemma \ref{lemma:activeIsWritten} and the \emph{passiveCollide} function it follows that a passive colliding multi-op executes \lref{FINISHCOL:spin} only if the active colliding multi-op it collided with has the same semantics.
\end{proofsketch}

\begin{definition}
\label{def:signaling}
We say that a multi-op is a \emph{delegate multi-op} if it is not a waiting multi-op.\\
We call the singly linked list of the multiOp structures that starts in the multiOp of the delegate multi-op, the \emph{multiOp list} of the delegate multi-op.
If $op_1$ is a delegate multi-op and $op_2$ is another multi-op whose multiOp structure is in the multiOp list of $op_1$ then we say that $op_2$ is in the multiOp list of $op_1$.\\
If the delegate multi-op is push, we call the singly linked list of cells that starts with the cell of the delegate operation the \emph{cells list} of the delegate operation.
\end{definition}

\begin{lemma}
\label{lemma:allTheSameType}
Let $op_1$ be a delegate multi-op, then all multi-ops in its multiOp list except for $op_1$ itself are waiting multi-ops and all the multi-ops in the multiOp list of $op_1$ have identical operation semantics.
\end{lemma}

\begin{proofsketch}
Let $op_1$ be a delegate multi-op.
Upon the first invocation of $op_1$, the only multi-op in $op_1$'s multiOp list is $op_1$ so the claim holds.
In addition, note that the last field in $op_1$'s multiOp points to itself, which is the last multiOp in $op_1$'s multiOp list.\\
Multi-ops are added to the multiOp list of $op_1$ only during the execution of the \texttt{combine} function which is called during the execution of \texttt{activeCollide} and only if some other $op_2$ is a passive colliding multi-op that collided with $op_1$ and has identical semantics.
According to Lemma \ref{lemma:waitingIsCombining}, a waiting multi-op is a passive colliding multi-op with the same operation semantics as the active colliding multi-op it collided with.
Let $op_2$ denote the waiting multi-op that is added to $op_1$'s multiOp list.
Note that, $op_1$ adds $op_2$'s multiOp to its multiOp list at its end and then updates the last field of its multiOp structure to point to the last element in $op_2$'s multiOp list.
As $op_2$ is a waiting multi-op with the same operation semantics as $op_1$, the claim holds.
\end{proofsketch}

\begin{definition}
\label{def:signaling}
Let $op_1$ and $op_2$ be two multi-ops. We say that $op_1$ \emph{signals} SIGNAL to $op_2$ if $op_1$ writes SIGNAL to the \emph{cStatus} field of the multiOp structure of $op_2$.
\end{definition}

\begin{definition}
\label{def:CombinerOf}
Let $op_1$ and $op_2$ be two multi-ops. We say that $op_1$ is the delegate multi-op of $op_2$ in time $t$, if $op_1$ was not yet eliminated (in the \emph{multiEliminate} function) and one of the following cases holds:
\begin{enumerate}
\item $op_2$ is $op_1$ (that is, every delegate operation is the delegate operation of itself).
\item $op_1$ is an active colliding multi-op, $op_2$ is the passive colliding multi-op it collided with and $op_1$ and $op_2$ have the same operation semantics. In this case, $op_1$ becomes the delegate multi-op of $op_2$ when they collide (at which time $op_2$ ceases to be the delegate multi-op of itself).
\item $op_3$ is the delegate multi-op of $op_2$ and, in time $t$, $op_1$ becomes the delegate multi-op of $op_3$. In this case, we say that $op_1$ becomes the delegate multi-op of $op_2$ instead of $op_3$ ($op_3$ ceases to be the delegate multi-op of $op_2$).
\item Assume that $op_3$ is the delegate multi-op of both $op_1$ and $op_2$, the multiOps of both of them are in the multiOp list of $op_3$ and the multiOp of $op_1$ appears before the multiOp of $op_2$ in the multiOp list of $op_3$. If $op_3$ signals RETRY to $op_1$, then $op_1$ becomes the delegate multi-op of $op_2$. In this case we say that $op_3$ ceases being the delegate multi-op of $op_2$ and that $op_1$ starts being the delegate multi-op of $op_2$ (note that in this case $op_1$ stops executing \lref{FINISHCOL:spin} and by Definition \ref{def:waitingOperations} ceases being a waiting multi-op and starts being a delegate multi-op).
\end{enumerate}
\end{definition}

%%%%%%%%%%%%%%%%%%%%%%%%%%%%%%%%%%%%%%%%%%%%%%%%%%%%%%%%%%%%%%%%%%%%%%%%%%%%%%%%%%%%%%%%%%%%%%%%%%%%%%%%%%%%%%%%%%%%%%%%%%%%%%%%%%%%%%%%%%%%%%%%%%%%%%%%%%%%
%delegate operation lemmas:

\begin{lemma}
\label{lemma:exactly1Combiner}
Every waiting multi-op has exactly a single delegate multi-op.
\end{lemma}
\begin{proofsketch}
Let $op_2$ be a waiting multi-op. By Lemma \ref{lemma:waitingIsCombining}, $op_2$ is a passive colliding multi-op that collided with an active colliding multi-op with identical semantics, which we denote as $op_1$.
By Definition \ref{def:CombinerOf}, $op_1$ is a delegate multi-op of $op_2$ at the time both of them collide. Also from Definition \ref{def:CombinerOf}, a multi-op ceases to be the delegate multi-op of $op_2$ only if it is replaced by a single new delegate multi-op.

%. Thus a waiting multi-op has at least a single delegate operation.
%Assume towards a contradiction that $op_1$ is the delegate multi-op of $op_2$ and another multi-op $op_3$ becomes the %delegate multi-op of $op_2$ in addition to $op_1$.
%By definition \ref{def:CombinerOf}, this could only happen if $op_3$ is an active colliding multi-op colliding with %$op_2$ as the passive colliding multi-op.
%However, because $op_1$ is already an active colliding multi-op that collided with $op_2$ that's a contradiction to %lemma \ref{lemma:collidesWithOne}.
\end{proofsketch}

\begin{lemma}
\label{lemma:updatedCombiner}
Let $op$ be a delegate multi-op and let \emph{opInfo} be its multiOp structure, then the following properties hold when $op$ begins executing \texttt{cMultiPop}, \texttt{cMultiPush} or \texttt{collide}:
\begin{itemize}
\item The multiOp list of $op$ is comprised of $op$'s multiOp and of its waiting multi-ops.
\item \emph{opInfo}.length is the length of the multiOp list of $op$.
\item \emph{opInfo}.last points to the last multiOp in $op$'s multiOp.
\item If $op$ is a push operation, then the order of its cells list corresponds to its multiOp list.
\end{itemize}
\end{lemma}
\begin{proofsketch}
Upon an invocation of $op$, it is a delegate multi-op with a single multi-op ($op$ itself).
The multiOp's length field is initialized to 1, and last points to $op$'s multiOp.
If the multi-op is a push, its cell field contains the data with which $op$ was invoked and the cell's next field points to \Null. We still need to consider the following two cases in which the multiOp list is modified:
\begin{enumerate}
\item $op$ is the aInf parameter of the \texttt{combine} function.
\item $op$'s multiOp list was set on \llref{ELIMINATE:aCurrLength}{ELIMINATE:aCurrRetry} or \llref{ELIMINATE:pCurrLength}{ELIMINATE:pCurrRetry} of \texttt{multiEliminate} function.
\end{enumerate}
In the first case, $op$ is the active colliding multi-op which collides with some other passive colliding multi-op $op'$.
The \texttt{combine} function sets the next element of the last element in $op$'s multiOp list to point to $op'$ multiOp list's first element. It then sets $op$'s last pointer to point to the last element in $op'$'s multiOp list, and adds the number of entries of $op'$'s multiOp list to the length field of $op$.
In case both operations are of a push semantics, the function concatenates $op'$'s cells list to the end of $op$'s cells list.
%adi{Should this be done by induction?}\\
In the second case, $op$ is a delegate multi-op because some other operation, $op''$, signaled $op$ with RETRY in \texttt{multiEliminate}.
Prior to that signal, the last and length fields of $op$ were updated by $op''$.
The last field was set to $op''$'s last element in its multiOp list, and length was set to the number of multi-ops which were not eliminated from $op''$ multiOp list.
\end{proofsketch}

%%%%%%%%%%%%%%%%%%%%%%%%%%%%%%%%%%%%%%%%%%%%%%%%%%%%%%%%%%%%%%%%%%%%%%%%%%%%%%%%%%%%%%%%%%%%%%%%%%%%%%%%%%%%%%%%%%%%%%%%%%%%%%%%%%%%%%%%%%%%%%%%%%%%%%%%%%%%

\begin{lemma}
\label{lemma:correctPush}
Let $op$ be a multi-op returning \True in \lref{CPUSH:returnTrue} of \texttt{cMultiPush}, then the following properties hold:
\begin{enumerate}
\item Each multi-op in $op$'s multiOp list is a waiting push multi-op except for the first multi-op which is the delegate's multi-op.
\item Before returning \True on \lref{CPUSH:returnTrue}, $op$ signaled FINISHED to each waiting multi-op in its multiOp list.
\item All cells in the cells list of $op$ were pushed to the central stack in the reverse order of the cells list.
\end{enumerate}
\end{lemma}
\begin{proofsketch}
\begin{enumerate}
\item If $op$ returns \True in \lref{CPUSH:returnTrue} then $op$ is a push multi-op and, from Lemma \ref{lemma:allTheSameType}, all the multi-ops in the delegate multiOp list are waiting push multi-ops except for the first multi-op, which is the delegate multi-op.
\item Clear from the code of the \texttt{cMultiPush} function.
\item From Lemma \ref{lemma:updatedCombiner}, the length field of $op$'s multiOp is equal to the length of $op$'s multiOp list, and the order of the cells list corresponds to the multiOp list. From the code on \llref{CPUSH:setlast}{CPUSH:castop}, after a successful CAS operation, CentralStack points to the first cell of $op$'s cells list, and the last cell of $op$'s cells list points to the previous top item of the central stack. It is easy to see that the reversed order of the cells list implicates an execution of $op$'s multiOp list push multi-ops starting from its last multi-op and ending with the delegate multi-op.
\end{enumerate}
\end{proofsketch}

\begin{lemma}
\label{lemma:correctPop}
Let $op$ be a delegate multi-op that returns \True in \lref{CPOP:ReturnEmpty} or \lref{CPOP:ReturnTrue} of \texttt{cMultiPop}, then the following properties hold:
\begin{enumerate}
\item Each multi-op in $op$'s multiOp list is a waiting pop multi-op except for the first multi-op which is the delegate multi-op.
\item Before returning \True on \lref{CPOP:ReturnEmpty} or \lref{CPOP:ReturnTrue}, $op$ signals FINISHED to each waiting multi-op in its multiOp list.
\item The number of cells $op$ has popped from the central stack in \texttt{cMultiPop} is equal to the minimum between the number of multiOp elements in $op$'s multiOp list and the number of cells in the stack.
\item If $op$ writes EMPTY\_CELL in the cell field of some multiOp in its multiOp list, then the number of elements in the central stack was smaller than the number of elements in $op$'s multiOp list.
\item If $op$ writes a cell in the cell field of a multiOp from its multiOp list, then that cell was previously pushed to the stack by some push operation and that multiOp is the only multiOp of a pop operation pointing to that cell.
\end{enumerate}
\end{lemma}
\begin{proofsketch}
\begin{enumerate}
\item If $op$ returns \True in \lref{CPOP:ReturnEmpty} or \lref{CPOP:ReturnTrue} then $op$ is a pop multi-op and from Lemma \ref{lemma:allTheSameType} all the multi-ops in the delegate multiOp list are waiting pop multi-ops  except the first multi-op which is the delegate multi-op.
\item Clear from the code of the \texttt{cMultiPop} function.
\item From Lemma \ref{lemma:updatedCombiner}, the length field of the multiOp of $op$ is the length of the multiOp list of $op$. Assume that the length of the multiOp list of $op_1$ is $N$. If the stack is not empty, on \lref{CPOP:IfPopCAS} CentralStack points to the $N + 1$'th cell in the stack and if the stack has $N$ items or less, CentralStack points to \Null.
\item Immediate from the code on \llref{CPOP:BeginCheckNextLoop}{CPOP:EndCheckNextLoop}.
\item \Llref{CPOP:PoppingTinfo}{CPOP:EndCheckNextLoop} imply that each cell that is assigned to a multi-op is a cell which was taken from the central stack. According to Lemma \ref{lemma:correctPush}, all cells in the central stack were inserted by a successful push multi-op using \texttt{cMultiPush}.
\end{enumerate}
\end{proofsketch}

\begin{lemma}
\label{lemma:pushAtMostOnePop}
Let $op$ be a delegate push multi-op returning \True in \lref{CPUSH:returnTrue} of \texttt{cMultiPush}, then each of the cells it inserts to the central stack will be assigned to at most a single pop operation.
\end{lemma}
\begin{proofsketch}
Assume by way of contradiction that more than two pop operations, $op_1$ and $op_2$, were assigned the same cell which was pushed to the central stack.
Without loss of generality we assume that $op_1$ performed the successful CAS on \lref{CPOP:IfPopCAS} before $op_2$.
According to Lemma \ref{lemma:correctPop} and the code, $op_1$ assigns this cell to one of the operations in its multiOp list, and accordingly CentralStack then points to a cell which is not one of the removed cells.
When $op_2$ performs the successful CAS on \lref{CPOP:IfPopCAS}, all the cells removed by $op_2$ are different from the cells removed by $op_1$, in contradiction to the assumption.
\end{proofsketch}

\begin{definition}
\label{def:eliminatingOperation}
We say that an operation is an \emph{eliminated} multi-op if it returns \True on \lref{STARTCOL:returnEliminate} or on \lref{FINISHCOL:returnEliminate} or some other thread signals it FINISHED on \lref{ELIMINATE:activeOpFinished} or on \lref{ELIMINATE:passiveOpFinished}.
\end{definition}

\begin{lemma}
\label{lemma:correctElimination}
Let $op$ be an eliminated multi-op, then the following properties hold:
\begin{itemize}
\item If $op$ is a pop, it obtains the value of a single eliminated push operation.
\item If $op$ is a push, its value is obtained by a single eliminated pop operation.
\end{itemize}
\end{lemma}
\begin{proofsketch}
Let $op_1$ and $op_2$ be two delegate multi-ops where WLOG $op_1$ is an active colliding multi-op and $op_2$ is the passive colliding multi-op it collided with, and the two multi-ops have reverse operation semantics.
We will begin by proving the lemma for the delegate operations themselves.
If $op_1$ returns \True on \lref{STARTCOL:returnEliminate}, then it is after it executed \texttt{multiEliminate} between multiOps' list of $op_1$ and $op_2$.
In case $op_1$ is a push multi-op, $op_2$ executes \texttt{passiveCollide} and obtains the cell of $op_1$ on \lref{FINISHCOL:setCell}.
If, however, $op_1$ is a pop multi-op, it obtains the cell of $op_2$ while executing the first iteration on \llref{ELIMINATE:activePop}{ELIMINATE:endIf}.
Note that $op_2$ will return \True on \lref{FINISHCOL:returnEliminate}.
\\
We now prove the elimination of the waiting multi-ops of $op_1$ and $op_2$.
Let $op_1'$ be the $M$'th operation in $op_1$'s multiOp list and $op_2'$ be the $M$'th operation in $op_2$'s multiOp list.
If $op_1'$ is a push operation then its value is obtained by $op_2'$ on \lref{ELIMINATE:matchCell2}.
However, if $op_1'$ is a pop operation then it obtains $op_2'$'s value on \lref{ELIMINATE:matchCell1}.
Note that $op_1$ then signals both $op_1'$ and $op_2'$ FINISHED on \llref{ELIMINATE:activeOpFinished}{ELIMINATE:passiveOpFinished}.

\end{proofsketch}

\begin{lemma}
\label{lemma:correctCombining}
Let $op_2$ be an operation returning \True in \lref{FINISHCOL:returnFinished} and let $op_1$ be the operation that signaled FINISHED to $op_2$, then:
\begin{itemize}
\item If $op_1$ is a pop operation then it obtains the value of a push operation or EMPTY\_CELL if the stack was empty when its operation was performed.
\item If $op_1$ is a push operation then at most a single pop operation returns its item.\\
\end{itemize}
\end{lemma}
\begin{proofsketch}
Clearly from the code, $op_1$ could have signaled $op_2$ FINISHED only in \lref{CPOP:ReleaseAll}, \lref{CPOP:ReleasingTinfoNext}, \lref{CPUSH:setFinish}, \lref{ELIMINATE:activeOpFinished} or \lref{ELIMINATE:passiveOpFinished}.
If $op_1$ signaled FINISHED in \lref{CPOP:ReleaseAll} or \lref{CPOP:ReleasingTinfoNext}, then $op_1$ is a pop operation bound to return \True in \lref{CPOP:ReturnEmpty} or \lref{CPOP:ReturnTrue} and by Lemma \ref{lemma:correctPop} it follows that $op_2$ is a pop operation obtaining the value of a single push operation.\\
If $op_1$ signaled FINISHED in \lref{CPUSH:setFinish}, than $op_1$ is a push operation bound to return \True in \lref{CPUSH:returnTrue}. By Lemma \ref{lemma:allTheSameType} $op_2$ is also a push operation and by Lemma \ref{lemma:pushAtMostOnePop} its item will be returned by at most one pop operation.
Finally, if $op_1$ signals $op_2$ FINISHED in \lref{ELIMINATE:activeOpFinished} or \lref{ELIMINATE:passiveOpFinished}, then by Lemma \ref{lemma:correctElimination} the proof holds.

\end{proofsketch}

\begin{theorem}
\label{def:correctPoolSemantics}
The Dynamic Elimination-Combining Stack has correct pool semantics.
\end{theorem}
\begin{proofsketch}
A thread can finish its operation by returning \True in one of the following lines: \lref{CPOP:ReturnEmpty}, \lref{CPOP:ReturnTrue}, \lref{CPUSH:returnTrue}, \lref{COLLIDE:caslocation1success}, \lref{COLLIDE:caslocation1fail} or \lref{COLLIDE:caslocation2fail}.
If a thread finishes its operation in \lref{CPOP:ReturnEmpty} or \lref{CPOP:ReturnTrue}, then it is a pop operation performed on the central stack and, by Lemma \ref{lemma:correctPop}, the operation is a correct pool operation.
If a thread finishes its operation in \lref{CPUSH:returnTrue}, then it is a push operation performed on the central stack and, by Lemma \ref{lemma:pushAtMostOnePop}, the operation is a correct pool operation.

If a thread finishes its operation in \lref{COLLIDE:caslocation1success}, then it is an active colliding multi-op that performed an elimination and, by Lemma \ref{lemma:correctElimination}, the operation is a correct pool operation.
If a thread finishes its operation in \lref{COLLIDE:caslocation1fail} or \lref{COLLIDE:caslocation2fail}, then it is a passive colliding multi-op that returned \True either from \lref{FINISHCOL:returnEliminate} or from \lref{FINISHCOL:returnFinished} in \texttt{passiveCollide}.
If the thread return \True in \lref{FINISHCOL:returnEliminate} then its operation was performed by  elimination and, by Lemma \ref{lemma:correctElimination}, it is a correct pool operation. On the other hand, if the thread returned \True from \lref{FINISHCOL:returnFinished} then its operation was performed by combining and, by Lemma \ref{lemma:correctCombining}, it is also a correct pool operation. It follows from Definition \ref{def:pool} that DECS has correct pool semantics.
\end{proofsketch}
%%%%%%%%%%%%%%%%%%%%%%%%%%%%%%%%%%%%%%%%%%%%%%%%%%%%%%%%%%%%%%%%%%%%%%%%%%%%%%%%%%%%%%%%%%%%%%%%%%%%%%%%%%%%%%%%%%%%%%%%%%%%%%%%%%%%%%%%%%%%%%%%%%%%%%%%%%%%%

\subsection*{Linearizability}

\begin{lemma}
\label{lemma:LinPoints}
The linearization points of the DECS algorithm are as follows:\\
All active operations are linearized in the following lines, executed in their last iteration of the \texttt{push} (\llref{MAIN:pushcentral}{MAIN:pushcollideret}) or \texttt{pop} (\llref{MAIN:popcentral}{MAIN:popcollideret}) operation:
\begin{itemize}
\item Delegate thread pop operations are linearized in \lref{CPOP:CheckTop}, \lref{CPOP:IfPopCAS} and in \lref{STARTCOL:cas} when the collision is an eliminating collision.
\item Delegate thread push operations are linearized in \lref{CPUSH:castop} and \lref{STARTCOL:cas} when the collision is an eliminating collision.
\end{itemize}
Eliminating passive operations are linearized in the linearization point of their single matching active colliding multi-op, where push colliding multi-op are linearized before the pop operation.\\
Waiting operations (combining passive operations) are linearized in the linearization point of their single delegate operation, where:
\begin{itemize}
\item If their delegate operation is a pop operation and its linearization point is \lref{CPOP:CheckTop} or \lref{CPOP:IfPopCAS}, then all the corresponding waiting operations are linearized at that point, according to their order in the multiOp list.
\item If their delegate operation is a push operation and its linearization point is \lref{CPUSH:castop}, then all the corresponding waiting operation are linearized at that point, in \emph{reverse} order of the multiOp list.
\item If their delegate operation linearization point is \lref{STARTCOL:cas}, then the linearization point of each waiting operation occurs at the linearization point of corresponding delegate operation and push colliding multi-op are linearized before pop colliding multi-ops.
\end{itemize}
\end{lemma}

\begin{proofsketch}
\Lref{CPOP:CheckTop}, \lref{CPOP:IfPopCAS}, \lref{CPUSH:castop} and \lref{STARTCOL:cas} complete by modifying the central stack; in this case, the claim follows by a simple extension of the Treiber \cite{treiber} algorithm correctness argument for multi-ops.

Next, we consider the linearization points for passive eliminating operations. In order to prove the claim for these operations, we need to prove that the linearization as defined in the lemma statement occurs during the operation's time interval. Let $op_2$ be a passive colliding multi-op and let $op_1$ be its matching active colliding multi-op. Assume by way of contradiction that $op_2$ terminated before the linearization point defined in the lemma statement. From Definitions \ref{def:collidingOperation} and \ref{def:eliminatingOperation}, $op_2$ executed \llref{FINISHCOL:getInfo}{FINISHCOL:returnEliminate}. Specifically, $op_2$ writes \Null to its entry in the location array and finishes its operation whereas $op_1$ performs \lref{STARTCOL:cas} only later. From the condition of \lref{COLLIDE:checkinfo}, it follows that the multiOp of $op_2$ must be non-\Null for $op_1$ to succeed in colliding with it. It follows that $op_1$ fails in the CAS performed in \lref{STARTCOL:cas}, in contradiction to Definition \ref{def:collidingOperation}.

We now consider the linearization points defined for waiting operations. By Lemma \ref{lemma:exactly1Combiner}, every waiting operation has exactly one delegate operation at any point of time, thus the linearization points are well-defined. By the definition of waiting operations, these operation did not yet terminate at their linearization times.

Finally, we consider the linearization points for operations that complete by elimination (these may be active colliding operations, passive colliding operations, or waiting operations). We already showed above that the linearization points for passive colliding operations and waiting operations are well defined. We are left to prove that correct LIFO ordering is maintained between any two linearized colliding operations and between these operations and operations that complete by modifying the central stack.

When linearizing a passive collider in the linearization point of its single active matching operation, no other operations can be linearized between these two operations. Moreover, since the push operation is linearized immediately before the pop operation, this is a legal LIFO ordering that cannot interfere with LIFO matchings of other collisions or with operations completed on the central stack. From Lemma \ref{lemma:correctElimination}, a pop operation obtains the value of the push operation it collides with.

Similar arguments establish the linearizability of the waiting operations in the combining lists of a pair of delegate operations that eliminate with each other. In this case, from Lemma \ref{lemma:correctCombining}, a waiting pop operation returns the value it obtained from the matching push operation.
\end{proofsketch}

\begin{theorem}
\label{def:LinPointsCorrect}
The Dynamic Elimination-Combining Stack is a correct linearizable implementation of a stack object.
\end{theorem}
\begin{proofsketch}
Immediate from Lemma \ref{lemma:LinPoints}.

\end{proofsketch}

\subsection*{Deadlock Freedom}
\begin{theorem}
\label{def:deadlockFreedom}
The dynamic elimination-Combining stack algorithm is deadlock-free.
\end{theorem}
\begin{proofsketch}
Let $op$ be an operation. We show that if each thread is scheduled infinitely often, then in every iteration made by $op$, some operation is linearized.

If $op$ is an active delegate operation, then it first tries to apply the operations in its multiOp list (which includes its own operation) to the central stack. If $op$ succeeds, then its linearization point has occurred. Otherwise, it must be that $op$ fails the CAS in \lref{CPOP:IfPopCAS} or \lref{CPUSH:castop} and this can only occur if another operation applied a successful CAS -- and was therefore linearized -- in the course of $op$'s iteration.

If $op$ is an active or a passive eliminating operation that succeeds in colliding, then its linearization point has already occurred and in \lref{FINISHCOL:returnEliminate} it returns \True and finishes its operation.

Otherwise, $op$ is a waiting operation. Let $op_1$ be the single delegate multi-op such that when $op_1$ is linearized, it is the delegate multi-op of $op$. From Lemmma \ref{lemma:LinPoints}, $op$ is linearized at the same time as $op_1$. More specifically, the linearization point of $op_1$ is followed by a release of all the waiting operations in its multiOp list. This is done by signaling them a FINISHED value. From Lemma \ref{lemma:updatedCombiner}, when $op_1$ is linearized, $op$ is in its multiOp list. Thus, as $op_1$ continues taking steps after performing its linearization point, $op$ is eventually released from its waiting at \lref{FINISHCOL:spin} and finishes its operation. To conclude the proof, observe that in every iteration of $op$, either $op$'s delegate thread or another delegate thread is linearized and will then release all of its waiting operation.
\end{proofsketch}

\section*{Appendix C: The Non-Blocking DECS algorithm}
\label{sec:nbdecs}

\SetKw{BoundedAwait}{bounded-await}

In Section \ref{sec:NB-DECS}, we provided a high-level description of the NB-DECS algorithm. In this section we provide a more detailed description of the differences in the pseudo-codes of DECS and NB-DECS.
We only describe the modifications that need to be incorporated into the blocking DECS algorithm presented in Section \ref{sec:algorithm}.

In order to highlight the differences between the two algorithms, we marked modified or new pseudo-code lines with red color, and kept the original code written in black color. The headers of new functions and functions that were extensively modified are also marked by red color.

New structure fields added for NB-DECS are shown in Figure \ref{fig:NBStructures}. An \emph{invalid} flag was added to each \emph{Cell}, indicating whether or not the cell was invalidated by a thread that stopped waiting.
Two fields were added to to the \texttt{multiOp} structure: (1) \emph{other} - which points to a multiOp structure of a thread with reverse semantics to indicate that an elimination between the two operations occurred; and (2) \emph{invalid} - indicating whether the multiOp structure is still valid (was not invalidated by the waiting thread).
\emph{invalid} flags are set using test-and-set to avoid race conditions.

A key change as compared to DECS is that a waiting thread (i.e., a thread that delegates its operation to another thread) stops waiting for its operation to be executed after some period of time.
That is, a waiting thread ``gives-up'' waiting for the delegate thread and retries its operation.

Each thread invokes a \emph{push} or \emph{pop} operation by executing the \texttt{push} and \texttt{pop} functions respectively (Figure \ref{fig:nbmainops}). As in DECS, the threads iteratively attempt to apply their operation on the \emph{CentralStack} structure and, if they fail, they access the elimination-combining layer.
After accessing the elimination-combining layer, each thread verifies that its multiOp is still valid to use.
If the multiOp is marked as invalid (and therefore cannot be used), the thread initializes a new multiOp record and re-attempts its operation on the \emph{CentralStack}.

\begin{figure}[b!]
\caption[Data Structure]{Non Blocking DECS Data Structures - modifications}
\label{fig:NBStructures}

%\NoCaptionOfAlgo
\nocaptionofalgo
\begin{algorithm}[H]
%\SetAlgoLined
\SetLine
%%%%%%%%%%%%%%%%%%%%%%%%%%%%%%%%%%%%%%%%
%\caption{\textbf{Figure 1}: Data Structures and Shared Variables}
%%%%%%%%%%%%%%%%%%%%%%%%%%%%%%%%%%%%%%%%
\Define  Cell: struct \{ \emph{data}: \Data, \emph{next}: \Cell, \\
\color{red}
\hspace{3.05cm} \emph{invalid}: \Bool \Init \False\}\;
\color{black}
\Define  multiOp: struct \{ \emph{id}, \emph{op}, \emph{length}, \emph{cStatus}: \Int, \emph{cell}: \Cell, \emph{next}, \emph{last}: \multiOp;\\
\color{red}
\hspace{3.05cm}  \emph{other}: \multiOp \Init \Null, \emph{invalid}: \Bool \Init \False\}\;
\color{black}
\end{algorithm}
\end{figure}

\begin{figure}[t!]
\caption[Native Operations]{Non Blocking push and pop operations}
\label{fig:nbmainops}
\begin{multicols}{2}

%\NoCaptionOfAlgo
\nocaptionofalgo
\begin{algorithm}[H]
%\SetAlgoLined
\SetLine
%%%%%%%%%%%%%%%%%%%%%%%%%%%%%%%%%%%%%%%%
\caption{\textbf{Data} \texttt{pop}()}
%%%%%%%%%%%%%%%%%%%%%%%%%%%%%%%%%%%%%%%%
    \multiOp mOp = initMultiOp()\nllabel{NBMAIN:popinit}\;
    \While{\True}{
        \uIf{cMultiPop(mOp) \nllabel{NBMAIN:popcentral}}{
            \Return mOp.cell.data\nllabel{NBMAIN:popcentralret}\;
        }\uElseIf{collide(mOp) \nllabel{NBMAIN:popcollide}}{
            \Return mOp.cell.data\nllabel{NBMAIN:popcollideret}\;
        }
\color{red}
		\ElseIf{mOp.invalid=\True}{mOp = initMultiOp()\;}
\color{black}
    }
\end{algorithm}

%\NoCaptionOfAlgo
\nocaptionofalgo
\begin{algorithm}[H]
%\SetAlgoLined
\SetLine
%%%%%%%%%%%%%%%%%%%%%%%%%%%%%%%%%%%%%%%%
\caption{\texttt{push}(\textit{Data}: data)}
%%%%%%%%%%%%%%%%%%%%%%%%%%%%%%%%%%%%%%%%
    \multiOp mOp = initMultiOp(data)\nllabel{NBMAIN:pushinit}\;
    \While{\True}{
        \uIf{cMultiPush(mOp) \nllabel{NBMAIN:pushcentral}}{
            \Return\;
        }\uElseIf{collide(mOp) \nllabel{NBMAIN:pushcollide}}{
            \Return\;
        }
\color{red}
		\ElseIf{mOp.invalid=\True}{mOp = initMultiOp(data)\;}
\color{black}
    }
\end{algorithm}
\end{multicols}
\end{figure}

\begin{figure}[t!]
\caption[Central Stack Operations]{Non-Blocking central stack operations}
\label{fig:NBMultiPop}

%\NoCaptionOfAlgo
\nocaptionofalgo
\begin{algorithm}[H]
%\SetAlgoLined
\SetLine
%%%%%%%%%%%%%%%%%%%%%%%%%%%%%%%%%%%%%%%%
\caption{\color{red}\textbf{boolean} \texttt{cMultiPop}(\textit{multiOp}: mOp)\color{black}}
%%%%%%%%%%%%%%%%%%%%%%%%%%%%%%%%%%%%%%%%
	\For{i = 1 \emph{\KwTo} mOp.length}{	
	    top = CentralStack\nllabel{NBCPOP:InitTop}\;
	    \If{top = \Null}{ \nllabel{NBCPOP:CheckTop}
	        \Repeat {mOp = \Null}
	        {\nllabel{NBCPOP:Release}
		        mOp.cell = EMPTY\_CELL\;
		        mOp.cStatus = FINISHED\nllabel{NBCPOP:ReleaseAll}\;
			testAndSet(mOp.invalid)\;
		        mOp = mOp.next\;
	        }
	        return \True\;\nllabel{NBCPOP:ReturnEmpty}
	    } \nllabel{NBCPOP:EndCheckTop}
	    next = top.next\nllabel{NBCPOP:InitNext}\;
	    \If{CAS(\&CentralStack, top, next)}{ \nllabel{NBCPOP:IfPopCAS}
			\If{testAndSet(top.invalid)=\False}{ \nllabel{NBCPOP:CheckValidTop}
				\If{delegatePop(mOp, top)=\False}{
					mOp.cell = top\nllabel{NBCPOP:SetDelegateCell}\;
					\Return \True\;\nllabel{NBCPOP:ReturnTrue}
				}
			}
		}
	}
	\Return \False\;
\end{algorithm}
\end{figure}

The code of the \texttt{cMultiPush} function is identical to that of the blocking version (Figure \ref{fig:DS-centralops-collide}-(c)). The code of \texttt{cMultiPop} changed, however, and is presented in figure \ref{fig:NBMultiPop}. A delegate thread which executes the \texttt{cMultiPop} operation attempts to execute $M$ (where $M$ is the length of its multiOp list) pop operations, one after the other.
If the stack is empty, the delegate thread iterates over its multiOp list, sets the \emph{cStatus} of the threads that are still waiting to FINISHED and their cell to an EMPTY\_CELL (\llref{NBCPOP:Release}{NBCPOP:ReturnEmpty}).
If the stack is non-empty, the delegate thread removes a single cell from the stack (\lref{NBCPOP:IfPopCAS}) and verifies that this cell is a valid cell (\lref{NBCPOP:CheckValidTop}).
When the delegate thread successfully removes a cell from the top of the stack, it searches for a thread in its multiOp list that is still waiting (using the \texttt{delegatePop} function). This is required, since, at least theoretically, it is possible that all waiting processes 'gave up' and stopped waiting.
If a waiting thread is not found, the removed cell is assigned to the delegate thread itself (\llref{NBCPOP:SetDelegateCell}{NBCPOP:ReturnTrue}).
If the CAS on \lref{NBCPOP:IfPopCAS} fails or the cell is invalid (\lref{NBCPOP:CheckValidTop}), the delegate thread continues and performs additional pop operation if required.
If no cell was obtained by the delegate thread for itself, the function returns \emph{false}, and the delegate thread enters the elimination-collision layer (as described in the \texttt{pop} function).

\begin{figure}[t!]
\caption[Pop Delegation]{A pop operation delegation assignment to a waiting thread}
\label{fig:delegatePop}
%\NoCaptionOfAlgo
\nocaptionofalgo
\begin{algorithm}[H]
%\SetAlgoLined
\SetLine
%%%%%%%%%%%%%%%%%%%%%%%%%%%%%%%%%%%%%%%%
\caption{\color{red}\textbf{boolean} \texttt{delegatePop}(\textit{multiOp}: mOp, \textit{Cell}: cell)\color{black}}
%%%%%%%%%%%%%%%%%%%%%%%%%%%%%%%%%%%%%%%%
	next = mOp.next\nllabel{DELEGATE:InitNext}\;
	\While{next $\neq$ \Null \nllabel{DELEGATE:StartWhile}}{
		next.cell = cell\nllabel{DELEGATE:SetCell}\;
		mOp.next = next.next\nllabel{DELEGATE:UpdateOpNext}\;
		\lIf{mOp.next = \Null}{mOp.last = mOp\nllabel{DELEGATE:UpdateOpLast}\;}
		mOp.length = mOp.length-1\nllabel{DELEGATE:UpdateOpLength}\;
		next.cStatus = FINISHED\nllabel{DELEGATE:SetFinish}\;
		\uIf{testAndSet(next.invalid) = \False \nllabel{DELEGATE:CheckValidOp}}{
			\Return \True\;\nllabel{DELEGATE:ReturnTrue}
		}\lElse{next = mOp.next\nllabel{DELEGATE:NextOp}}\;
	}
	\Return \False\;\nllabel{DELEAGTE:ReturnFalse}
\end{algorithm}
\end{figure}

The \texttt{delegatePop} function described in figure \ref{fig:delegatePop} receives a delegate thread's pop multiOp and a valid cell that was removed from the stack. The function searches for a waiting thread in the multiOp list and, if such a thread is found, assigns the cell to that waiting thread. The function starts by iterating over the multiOp list, one multiOp at a time, as long as no waiting multiOp is found. Once a waiting thread is found, the cell is assigned to it and the multiOp is removed from the delegate's multiOp list (\llref{DELEGATE:SetCell}{DELEGATE:UpdateOpLength}). In \lref{DELEGATE:CheckValidOp}, the waiting thread's multiOp is checked. If it is still valid (i.e., the thread is still waiting), the \texttt{delegatePop} function returns \emph{true}, indicating that a waiting thread which obtained the cell was found.
Otherwise, if no waiting thread is found, the function returns \emph{false}.

\begin{figure}[t!]
\caption[finishCollision]{Non blocking passive collide}
\label{fig:NBFinishCollision}
%\NoCaptionOfAlgo
\nocaptionofalgo
\begin{algorithm}[H]
%\SetAlgoLined
\SetLine
%%%%%%%%%%%%%%%%%%%%%%%%%%%%%%%%%%%%%%%%
\caption{\textbf{boolean} \texttt{passiveCollide}(\textit{multiOp}: pInf)}
%%%%%%%%%%%%%%%%%%%%%%%%%%%%%%%%%%%%%%%%
    aInf = location$[$pInf.id$]$\nllabel{NBFINISHCOL:getInfo}\;
    location$[$pInf.id$]$ = \Null\;\nllabel{NBFINISHCOL:setInfo}
    \eIf{pInf.op $\neq$ aInf.op \nllabel{NBFINISHCOL:compareOp}}{
        \If{pInf.op = POP \nllabel{NBFINISHCOL:isPop}}{
            pInf.cell = aInf.cell\nllabel{NBFINISHCOL:setCell}\;
        }
        \Return \True\;\nllabel{NBFINISHCOL:returnEliminate}
    }{\nllabel{NBFINISHCOL:beginSameSemantics}
\color{red}
        \BoundedAwait(pInf.cStatus $\neq$ INIT)\nllabel{NBFINISHCOL:awaitStart}\;
		\uIf {pInf.cStatus $\neq$ INIT}{
			\Switch{pInf.cStatus}{\nllabel{NBFINISHCOL:startSwitch}
				\uCase{FINISHED}{
					\Return \True\;\nllabel{NBFINISHCOL:returnFinished}
				}
				\uCase{EXCHANGE}{\nllabel{NBFINISHCOL:startEliminate}
					\uIf{testAndSet(pInf.other.invalid) = \False\nllabel{NBFINISHCOL:eliminateTas}}{
						\lIf{pInf.op = POP}{pInf.cell = pInf.other.cell\nllabel{NBFINISHCOL:exchangeCells}}\;
						\Return \True\;\nllabel{NBFINISHCOL:returnExchanged}
					}
					\Else{\nllabel{NBFINISHCOL:exchangeTasFail}
						initFields(pInf)\;
						\Return \False\;
					}
				}\nllabel{NBFINISHCOL:endEliminate}
				\Case{RETRY}{
					pInf.invalid = \False\;\nllabel{NBFINISHCOL:resetInvalid}
					pInf.cStatus = INIT\nllabel{NBFINISHCOL:setInit}\;
					\Return \False\;\nllabel{NBFINISHCOL:retryFalse}
				}
			}\nllabel{NBFINISHCOL:endSwitch}
		}\lElse{\nllabel{NBFINISHCOL:callWakeup}\Return wakeup(pInf)}\;
\color{black}
	}\nllabel{NBFINISHCOL:endSameSemantics}
\end{algorithm}
\end{figure}

As in DECS, the \texttt{passiveColide} function (presented in figure \ref{fig:NBFinishCollision}) is invoked by a passive collider after it identifies that it was collided with.
The passive collider first reads the multi-op pointer written to its entry in the \textit{location} array by the active collider and initializes its entry in preparation for future operations (\llref{NBFINISHCOL:getInfo}{NBFINISHCOL:setInfo}).
If the multi-ops of the colliding threads-pair are of reverse semantics (\lref{NBFINISHCOL:compareOp}) then the function returns \emph{true} in \lref{NBFINISHCOL:returnEliminate}\ because, in this case, it is guaranteed that the colliding delegate threads exchange values. Specifically, if the passive thread's multi-op type is \textit{pop}, the thread copies the cell communicated to it by the active collider (\lref{NBFINISHCOL:setCell}).

The algorithm in the non blocking version is different for the case where both multi-ops are of identical semantics (\llref{NBFINISHCOL:beginSameSemantics}{NBFINISHCOL:endSameSemantics}).
In this case, the passive collider's operations were delegated to the active thread until some time limit expires or the executing thread's \emph{cStatus} is modified, as shown in \lref{NBFINISHCOL:awaitStart}.
Upon terminating the bounded waiting, the passive collider's operation is either performed by the active thread or has yet to be performed. In the latter case, the \emph{cStatus} field of the passive operation is still INIT, and the function calls the \texttt{wakeup} function in \lref{NBFINISHCOL:callWakeup} so that the passive thread will retry executing its operation while avoiding a race condition with the active collider.

If, on the other hand, the active collider updated the \emph{cStatus} field, then the executing thread's operation is either: (1) executed on the central stack and finished, (2) eliminated with another operation, or (3) an eliminating operation was not found.
In the first case, the operation was successfully executed on the central stack by a thread that invoked the \texttt{cMultiPush} or the \texttt{cMultiPop} functions, and the function returns \emph{true} indicating the operation is terminated (\lref{NBFINISHCOL:returnFinished}).
The second case, where the operation was eliminated with another operation is dealt with in \llref{NBFINISHCOL:startEliminate}{NBFINISHCOL:endEliminate},
The executing thread invokes a test-and-set operation on the invalid flag of the operation that was assigned to it.
If the test-and-set succeeds, then the other operation is a valid operation, an elimination occurs and the function return \emph{true} in \lref{NBFINISHCOL:returnExchanged}. (Note that if the passive collider's operation is a pop, the passive collider obtains the cell of the other operation on \lref{NBFINISHCOL:exchangeCells}.)
If, on the other hand, the test-and-set fails -- implying that the other thread stopped waiting (\lref{NBFINISHCOL:exchangeTasFail}) -- the executing thread resets its multiOp record fields and returns \emph{false}. The third case, where \emph{cStatus} is set to RETRY, is similar to the corresponding case in the DECS algorithm: The passive operation is notified that an eliminating operation was not found, and so it clears its invalid flag, resets its \emph{cStatus} to INIT and returns \emph{false} indicating that the operation was not executed yet.

\begin{figure}[t!]
\caption[combineEliminate]{Non blocking elimination of multi operations}
\label{fig:NBMultiEliminate}

%\NoCaptionOfAlgo
\nocaptionofalgo
\begin{algorithm}[H]
%\SetAlgoLined
\SetLine
%%%%%%%%%%%%%%%%%%%%%%%%%%%%%%%%%%%%%%%%
\caption{\texttt{multiEliminate}(\textit{multiOp}: aInf, pInf)}
%%%%%%%%%%%%%%%%%%%%%%%%%%%%%%%%%%%%%%%%
    aCurr = aInf\nllabel{NBELIMINATE:setACurr}\;
    pCurr = pInf\;
    \Repeat{aCurr $\neq$ null \AndCond pCurr $\neq$ null}{
\color{red}
    aCurr.other = pCurr\;
    pCurr.other = aCurr\;
    aCurr.cStatus = EXCHANGE\;
    pCurr.cStatus = EXCHANGE\;
\color{black}
    aInf.length = aInf.length - 1\;
    pInf.length = pInf.length - 1\;
    aCurr = aCurr.next\;
    pCurr = pCurr.next\;
}\nllabel{NBELIMINATE:endMatching}
    \uIf{aCurr $\neq$ \Null\nllabel{NBELIMINATE:activeList}}{
\color{red}
	     \While{aCurr $\neq$ \Null}{
\color{black}
			aCurr.length = aInf.length\nllabel{NBELIMINATE:setACurrLength}\;
			aCurr.last = aInf.last\;
			aCurr.cStatus = RETRY\nllabel{NBELIMINATE:setACurrRetry}\;
\color{red}
			\lIf{testAndSet(aCurr.invalid)=\False}{\Return}\;\nllabel{NBELIMINATE:aCurrReturn}
			\Else{
				aInf.length = aInf.length - 1\nllabel{NBELIMINATE:aCurrLengthUpdate}\;
				aCurr = aCurr.next\nllabel{NBELIMINATE:aCurrNext}\;
			}
\color{black}
		}
  }\ElseIf{pCurr $\neq$ \Null\nllabel{NBELIMINATE:passiveList}}{
\color{red}
 	     \While{pCurr $\neq$ \Null}{
\color{black}
			pCurr.length = pInf.length\nllabel{NBELIMINATE:setPCurrLength}\;
			pCurr.last = pInf.last\;
			pCurr.cStatus = RETRY\nllabel{NBELIMINATE:setPCurrRetry}\;
\color{red}
			\lIf{testAndSet(pCurr.invalid)=\False}{\Return}\;\nllabel{NBELIMINATE:pCurrReturn}
			\Else{
				pInf.length = pInf.length - 1\nllabel{NBELIMINATE:pCurrLengthUpdate}\;
				pCurr = pCurr.next\nllabel{NBELIMINATE:pCurrNext}\;
			}
\color{black}
		}\nllabel{NBELIMINATE:passiveListEnd}
}
\end{algorithm}
\end{figure}

Figure \ref{fig:NBMultiEliminate} shows the \texttt{multiEliminate} function which is called by an active collider with two multiOp operations of reverse semantics.
As in DECS, the function iterates over the two lists of the given multi-ops on \llref{NBELIMINATE:setACurr}{NBELIMINATE:endMatching} until one of the lists has no more elements.
Note that contrary to the blocking \texttt{multiEliminate} function, the \emph{cStatus} of the operations is set to EXCHANGE (instead of FINISHED), and the cell is not obtained by the pop operation; instead, the \emph{other} field is set to the matching operation's multiOp record.
Setting the \emph{cStatus} to EXCHANGE allows passive threads, upon terminating their bounded await (\lref{NBFINISHCOL:awaitStart}), to observe that their operation was eliminated and that the cell of the push operation can be obtained by the pop operation.

After iterating over both lists, one of the lists may contain more elements (recall the ``residue'' of a multiOp list described in section \ref{sec:algorithm}).
If the active (passive) collider's multiOp list is longer than the list of the passive (active) collider, \llref{NBELIMINATE:activeList}{NBELIMINATE:passiveList} (\llref{NBELIMINATE:passiveList}{NBELIMINATE:passiveListEnd}) are executed on the residue list of the active (passive) collider.
The delegate thread executing the \texttt{multiEliminate} function iterates over the active (passive) residue list, searching for the first operation of a waiting thread, setting its \emph{cStatus} to RETRY after updating its multi-op list on \llref{NBELIMINATE:setPCurrLength}{NBELIMINATE:setPCurrRetry} (\llref{NBELIMINATE:setPCurrLength}{NBELIMINATE:setPCurrRetry}).
Contrary to the blocking algorithm, in this case a thread may no longer be waiting. 
To cope with such a scenario, after setting \emph{cStatus} to RETRY and updating the multiOp operations list, the delegate thread attempts to test-and-set the multiOp's invalid flag.
If the test-and-set operation succeeds, the multiOp is of a thread that is still waiting and the function returns on \lref{NBELIMINATE:aCurrReturn} (\lref{NBELIMINATE:pCurrReturn}).
If the test-and-set fails, then the thread has stopped waiting and resumed its operation, and the delegate thread updates the length of the multiOp list, and continues to the delegate thread's next multi-op of the active (passive) residue list on \llref{NBELIMINATE:aCurrLengthUpdate}{NBELIMINATE:aCurrNext} (\llref{NBELIMINATE:pCurrLengthUpdate}{NBELIMINATE:pCurrNext}).

\begin{figure}[t!]
\caption[wakeup]{A passive thread wakeup function}
\label{fig:wakeup}
%\NoCaptionOfAlgo
\nocaptionofalgo
\begin{algorithm}[H]
%\SetAlgoLined
\SetLine
%%%%%%%%%%%%%%%%%%%%%%%%%%%%%%%%%%%%%%%%
\caption{\color{red}\texttt{wakeup}(\textit{multiOp}: mOp)\color{black}}
%%%%%%%%%%%%%%%%%%%%%%%%%%%%%%%%%%%%%%%%
	\eIf{mOp.op = POP\nllabel{WAKEUP:IfPop}}{
		\uIf{testAndSet(mOp.invalid) = \True}{
			\eIf{mOp.cStatus = RETRY}{
				mOp.invalid = false\;
				mOp.cStatus = INIT\;
				\Return \False\;
			}{
				\uIf{mOp.cStatus $\neq$ null}{
					mOp.cell = mOp.other.cell\;
				}
				\Return \True\;
			}
		}\lElse{\Return \False}\;\nllabel{WAKEUP:PopTasFalse}
	}(\tcc*[f]{PUSH operation}){\nllabel{WAKEUP:IfPush}
		\uIf{testAndSet(mOp.cell.invalid) = \True \OrCond testAndSet(mOp.invalid) = \True\nllabel{WAKEUP:PushTas}}{
			\eIf{mOp.cStatus = RETRY\nllabel{WAKEUP:PushRetry}}{
				mOp.cStatus = INIT\nllabel{WAKEUP:PushSetINIT}\;
				mOp.invalid = \False\;
				\Return \False\;\nllabel{WAKEUP:ReturnRetry}
			}{\nllabel{WAKEUP:PushNotRetry}
				\Return \True\;\nllabel{WAKEUP:PushSuccess}
			}
		}\lElse{\Return \False}\;\nllabel{WAKEUP:PushTasFalse}
	}
\end{algorithm}
\end{figure}

The \texttt{wakeup} function presented in figure \ref{fig:wakeup} is invoked by a passive collider thread after it gives up waiting. The function is given a pointer to the passive collider's multiOp record and resolves a possible race condition with the active collider. If the operation is a pop (\llref{WAKEUP:IfPop}{WAKEUP:PopTasFalse}), the invalid flag of the operation is checked. If it is $0$, indicating that the active collider did not perform the passive collider's operation, the function returns \emph{false}. If the flag is set, however, then the \emph{cStatus} of the passive collider is now set to another value by the active thread, and the executing thread continues similarly to the code of \llref{NBFINISHCOL:startSwitch}{NBFINISHCOL:endSwitch} of \texttt{passiveCollide}.

\Llref{WAKEUP:IfPush}{WAKEUP:PushTasFalse} are executed in case the operation is a push.
In \lref{WAKEUP:PushTas}, two test-and-set operations are performed, one on the invalid flag of the passive operation's cell, and the other on the invalid flag of the operation structure itself.
If both invalid flags are $0$ (i.e., the cell and the multiOp record are both valid, and no other operation needs them), the function returns \emph{false} (\lref{WAKEUP:PushTasFalse}) indicating that the push operation should be restarted.
Note that the invalid flag of the cell can only be set by another pop operation, after popping the cell from the central stack (\lref{NBCPOP:CheckValidTop} in \texttt{cMultiPop}), and the operation's invalid flag is set when an elimination with another pop operation is done (\lref{NBFINISHCOL:eliminateTas} in \texttt{passiveCollide}).
In both cases, it is an indication that the push operation was executed (and its \emph{cStatus} was updated).
If the \emph{cStatus} field is set to RETRY, an elimination attempt was unsuccessful, and the executing thread resets its \emph{cStatus}, clears its invalid flag and returns \emph{false} to retry its push operation (\llref{WAKEUP:PushSetINIT}{WAKEUP:ReturnRetry}).
Otherwise (the value of \emph{cStatus} field is either EXCHANGE or FINISHED), the push operation was executed either by elimination or by inserting the element to the central stack, and \emph{true} is returned in  \lref{WAKEUP:PushSuccess}.

\end{document}